\documentclass[fleqn,usenatbib,useAMS]{mnras}

   \usepackage{graphicx}    
    \usepackage{amsmath}    
    \usepackage{amssymb} 
    \usepackage{pdflscape} 

    \newcommand{\kms}{\,km\,s$^{-1}$}
     
    \newcommand{\Msun}{\mbox{M$_{\odot}$}}

    \usepackage{newtxtext,newtxmath}
    \usepackage[utf8]{inputenc}
		
	\usepackage{color}

    \title[The nuclear architecture of NGC\,4151]{The nuclear architecture of NGC\,4151: on the path toward a universal outflow mechanism in light of NGC\,1068}
    \author[D. May et al.]{
    D. May $^{1}$\thanks{E-mail: dmay.astro@gmail.com},
    J.E. Steiner$^{1}$,
    R.B. Menezes$^{2}$,
    D.R.A. Williams$^{3}$,
    J. Wang$^{4}$
    \\
    $^{1}$Instituto de Astronomia, Geof\'isica e Ci\^encias Atmosf\'ericas, Universidade de S\~ao Paulo, 05508-090, S\~ao Paulo, SP, Brazil\\
    $^{2}$Centro de Ci\^encias Naturais e humanas, Universidade Federal do ABC, Santo Andr\'e, 09210-580 SP, Brazil\\
    $^{3}$ Department of Physics, University of Oxford, Denys Wilkinson Building, Keble Road, Oxford, OX1 3RH, UK\\
    $^{4}$ Department of Astronomy, Xiamen University, Fujian, 361005, China
	}
    \begin{document}
    
    \date{Draft for internal use only}

    \pagerange{\pageref{firstpage}--\pageref{lastpage}} \pubyear{2020}

    \maketitle

    \label{firstpage}

    \begin{abstract}
    
	We report near-infrared (NIR) IFU observations of the active galactic nucleus NGC\,4151 with archive data from the NIFS-Gemini North Telescope.
	We have selected best-seeing observations ($\lesssim$0.3 arcsec) that, allied to our methodology of image processing techniques, show structures with spatial resolution comparable to those of the HST.
	The intricate outflow of NGC\,4151 is revisited in light of the results found for NGC\,1068, in a previous work, and a very similar dynamic is found: the low-velocity [Fe\,{\sc ii}] emission depicts the glowing walls of an hourglass structure, while the high-velocity gas fills its volume. From this finding we show that the misalignment between the jet and the NLR is not a projection effect, as previously thought. A molecular outflow is detected for the first time in this galaxy and, just like in NGC\,1068, the transition between the molecular and the ionized gas phases comes from the fragmentation of molecular cavity walls into bullets of ionized gas exposed to the central source.
	Furthermore, it is suggestive that the same geometrical dichotomy between the cones seen in NGC\,1068 is found here, with one side, where the cavity is disrupted by the AGN, being more extended than the other. Finally, a new spatial correlation between the high-velocity [Fe\,{\sc ii}] and the soft X-ray emission of [Ne\,{\sc ix}] is found, which is unexpected given the difference between their ionization potentials.
	
    \end{abstract}


    \section{Introduction}
    \label{sec:intro}

  Outflows of ionized \citep{Fabian12,King15,Astor19} and molecular gas \citep{Davies14,Cicone14,Flu19} in active galactic nuclei (AGN) may be driven by plasma jets \citep{Wilson94,Koide99,Wu13} and/or radiation pressure \citep{Ciotti10,Crenshaw12,Faucher12,Tombesi12,Parker17}, both originating in the vicinity of an accretion disc feeding a supermassive black hole (SMBH). These mechanisms constitute important channels of galactic feedback, ejecting nuclear gas and accelerating the gas reservoir previously distributed in the interstellar medium (ISM) \citep{Pogge89,Das05,Das06,Dumas07,Muller11,Garcia14,Barbosa14}.
  The scales affected by the nuclear feedback range from the immediate surroundings of the SMBH \citep{Sadowsky13,Veilleux17} to kiloparsec distances into the intergalactic medium \citep{Gallimore06,Muk16}.
   
   The AGN ``big picture'' is described by the Unified Model, firstly proposed to explain the spectra of polarimetric observations of the nucleus of NGC\,1068 (e.g. \citealt{Antonucci85}). Although simple, the Unified Model is evoked to explain the presence or absence of broad permitted lines in the AGN spectra (namely, types 1 and 2, respectively), where the existence of both types of AGN is historically attributed to the viewing angle between an obscuring dusty torus (around the broad line region - BLR) and the line-of-sight. Such a torus is responsible to collimate the central radiation in form of two ionization cones \citep{Antonucci93,Urry95}, with the narrow line region (NLR) within their geometrical limits.
   
   However, there are multiple complexities seen in the morphologies, dynamics and ionization degrees inside the NLR of galaxies. With such a variety of phenomena, the current paradigm should be taken into account carefully and some questions may be raised: are the AGN tori of types 1 and 2 sources intrinsically similar \citep{Ramos11}? How to explain the high efficiency of the outflows in accelerating the gas along the NLR  \mbox{\citep{Crenshaw05}}? What are the implications of the frequent misalignment between the accretion disc and the torus \citep{Goosmann11,Fischer13,Fischer14}? How to explain the gas morphology that seems to preserve some symmetry along the NLR \citep{DMay17,Ardila17}?
   
   Therefore, analyzing outflows in nearby galaxies is the best way to spatially resolve their morphology down to small scales, pointing out possible connections between the gas dynamics, jets and the central source.
   With this approach, an extensive effort to study in detail the ionization cones of nearby galaxies have been done. We highlight some works that went deeper on the NLR architecture of some galaxies, like, in the optical, NGC\,1068 \citep{Evans91,Macchetto94,Cecil021068}, NGC\,4151 \citep{Evans93,Robinson94,Winge97,Kaiser00,Crenshaw15,Williams17}, NGC\,3393 \citep{Cooke00}, NGC\,3079 \citep{Cecil023079},  Mrk\,573 \citep{Revalski18b}, Mrk\,34 \citep{Revalski18}; and in the NIR, NGC\,1068 \citep{Muller09,Riffel14b,DMay17}, NGC\,4151 \citep{Thaisa094151,Thaisa104151}, Mrk\,1066 \citep{Rogemar10}, Mrk\,1157 \citep{Rogemar11}, Mrk\,79 \citep{Rogemar13}, NGC\,2110 \citep{Diniz15}, Mrk\,573 \citep{Fischer17}, NGC\,1386 \citep{Ardila17}.
   
   In reality, the NLR of galaxies does not always resemble a well-defined geometrical shape of two cones. The definition of ionization cones is just a volume that encompass the limits within which we normally detect the gas that is exposed to the collimated radiation from the AGN. Such approach is useful, for instance, to model the NLR kinematics and highlight its distribution in the geometrical context of the nuclear region \citep{Muller11,Fischer13}. There are cases, however, where the geometrical limits of a collimated radiation are remarkably visible in the form of the glowing walls of an hourglass, a well-known feature in some planetary nebulae - PNe (NGC\,6537 - \citealt{Cuesta95}, MyCn\,18 - \citealt{Sahai99}, Hubble\,12 - \citealt{Hora00}, NGC\,6302 - \citealt{S11}). In AGN, the hourglass shape of the NLR has been seen, for instance, in the integral emission of [Fe\,{\sc ii}] $\lambda$1.64 $\mu$m (NGC\,4151 - \citealt{Thaisa104151}, NGC\,1068 - \citealt{Riffel14b,Barbosa14} and NGC\,5728 - \citealt{Durre18}), in the H$_{2}$ $\lambda$2.12 $\mu$m (Mrk\,573 - \citealt{Fischer17}) and in some emission line images of NGC\,3393 (specially in the [S\,{\sc ii}]($\lambda 6717$+$\lambda6731$)/H$\alpha$ ratio - \citealt{Maksym17}). The presence of these symmetrical structures, at least in PNe, seems to be a natural consequence of gas ejection events, but there are still only a few of them detected in the NLR of AGN to draw some conclusion.
   
   However, a new result presented by \citet{DMay17} (hereafter M\&S17) for the galaxy NGC\,1068 re-ignited a particular interest in this hourglass shape for the NLR. These authors found a dichotomy between the low- and high-velocity [Fe\,{\sc ii}] emissions, where the low-velocity emission depicts the glowing walls of the hourglass and the high-velocity one is entirely contained in the inside. Since two of the most iconic and studied NLRs are probably in NGC\,1068 and NGC\,4151, this finding motivated us to look for the same behaviour in the latter. 
   Furthermore, thanks to the meticulous data treatment, M\&S17 have described a new connection between the dynamics of the ionized and molecular gas phases for NGC\,1068. There, the AGN is inflating a molecular bubble, which is being fragmented in compact clouds of ionized gas. Therefore, in this work we are not only looking for the [Fe\,{\sc ii}] dichotomy, but for a more complete view of the NLR of NGC\,4151 in light of the new results found for NGC\,1068, under the same methodology.  
   
   For this purpose, we use a set of IFU archival data of NGC\,4151 with the best seeing available ($\lesssim$0.3 arcsec), observed with adaptive optics (AO) and combined with  a thoughtful set of image processing techniques (better described in \citealt{Menezes14,Menezes15,Menezes19}). The wealthy amount of information of a data cube ($x,y,\lambda$) in the near-infrared (NIR), comprising emission lines with a wide range of ionization potentials (IP), allows us to carry out an almost ``artisanal'' work relating the distinct gas phases (like the molecular gas in the H$_{2}$ emission, the partially ionized gas with the [Fe\,{\sc ii}] lines, the ionized gas seen in the hydrogen recombination lines and the coronal line emission, here represented by the [Si\,{\sc vii}] $\lambda$2.48\,$\mu$m line) with their corresponding kinematics, at a high spatial resolution. 
   
   \subsection{NGC\,4151}
   \label{sec:n4151}
   
   As the brightest Seyfert 1 galaxy in the sky (also classified as Seyfert 1.5; \citealt{Osterbrock06}), NGC\,4151 hosts the most studied AGN of its class. We adopted the distance of 13.3 Mpc measured by \citet{Mundell03} (meaning that 1 arcsec is $\sim$65 pc for $H_{o}$=75 km $s^{-1}$ Mpc$^{-1}$), although a more recent work points out to a larger value of 19 Mpc \citep{Honig14}. NGC\,4151 has a morphological classification of (R\textquotesingle)SAB(rs)ab \citep{Vaucouleurs91} and, from the HI kinematics, an inclination of $i=21$\textdegree~and a position angle (PA) of 22\textdegree~\citep{Pedlar92} is inferred.
  
   One could say that NGC\,4151 was the first galaxy to defy the Unified Model \citep{Evans93} since it presents both a BLR - what defines a Seyfert 1 - and two clear ionization cones - expected in Seyfert 2. Such ``double behaviour'' means that the line-of-sight is very close to the edge of a clumpy torus, as firstly inferred by \citet{Robinson94} and, more recently, by \citet{Burtscher09} from infrared modeling.
   
   Beyond this peculiarity, the ionization cones of NGC\,4151 have some curious features, such as 1) the [O\,{\sc iii}] emission is highly filamentary and not directly associated with the radio jet \citep{Evans93,Mundell03,Williams17}; 2) the jet seems to have a negligible effect on the NLR kinematics \citep{Das05}, although \citet{Thaisa104151} advocate that the gas close to the systemic velocity is disturbed by the jet; 3) even though the jet seems to have a minor influence on the global NLR kinematics, there are some indications for the presence of shocks, as the origin of the low-luminosity clouds in back-flow seen by \citet{Das05}, some discrete high ratio of [Ne\,{\sc ix}]/[O\,{\sc vii}] close to the radio jet \citep{Wang11a} and that some local optical line ratios could be explained by gas compression due to the jet \citep{Winge97}; 4) if the misalignment between the jet and the ionization cones is, indeed, a projection effect (with the gas in the galactic disc intercepted by the cone's aperture (\citealt{Pedlar93,Robinson94}), then it is hard to explain the highest detected velocities (up to -1700 \kms - \citealt{Winge97,Kaiser00,Das05}) at distances $\gtrsim$50 pc from the nucleus. Such values would represent much higher deprojected velocities (v$\sim$3000 \kms) than those expected for rotation (or even outflows) in the galactic disc.
   
   This paper is organized as follows. In Section~\ref{sec:2}, we describe the observations, data reductions and treatment; in Section~\ref{sec:results}, we present the results, comparing the spatial morphology and resolution of the Br10 $\lambda$17367 \AA~and Br$\gamma$ $\lambda$21661 \AA~- after our data treatment - with the H$\alpha$ emission from the HST. We also show different [Fe\,{\sc ii}] velocity regimes, the complex molecular gas distribution and a concise analysis of the coronal and X-ray emissions. Then, in Section~\ref{sec:discussion}, we present our discussion about the origin of the ionized gas and the nature of the detected molecular outflow in light of the results found for NGC\,1068. Finally, we draw the conclusions of this work in Section~\ref{sec:conclusions}.

    \section{Observations}
    \label{sec:2}

    We used archival data from NIFS (Near-Infrared Integral-Field Spectrograph) \citep{McGregor03} at the Gemini North Telescope. NIFS operates with the AO module ALTAIR (ALTitude conjugate Adaptive optics for the InfraRed) with a pixel size of 0.103 arcsec~$\times$~0.043 arcsec and FoV of $\sim$3 arcsec~$\times$~3 arcsec. The spectral range comprises the $H$-band ($1.49-1.80~\mu$m), the $K$-band ($1.99-2.40~\mu$m) and the $K_{long}$ setting of the $K$ grating ($2.11-2.52~\mu$m), with a spectral resolution of 5290 ($\approx$ 3 \AA~ of spectral sampling or $\approx$60 km $s^{-1}$ in velocity). 
    The data were reduced using tasks of the NIFS package in IRAF environment. The procedure included trimming the images, flat-fielding, sky subtraction, correcting for spatial distortions and wavelength and flux calibrations. At the end of the process the data cubes were generated by the task \texttt{nifcube} with square spaxels of $\sim$0.05 arcsec~$\times$~0.05 arcsec.

    The three analyzed NIR gratings ($H$, $K$ and $K_{long}$) were all oriented at a PA=15\textdegree. The $H$-band observations were obtained by C. Onken \citep{Onken14}, under programme GN-2008A-Q-41, and the $K$-band and $K_{long}$-band by P. McGregor \citep{Thaisa104151}, under programme GN-2006B-C-9.

    The $H$-band data set is distinguished by the fact that we selected only observations with exceptional seeing, resulting in nine data cubes with seeing below 0.3 arcsec, with an average of 0.27 arcsec. We estimate the PSF of the $H$-band data from the standard star and nearly half of this value is found, that is, 0.138$\pm$0.005. To our knowledge, no emission lines from this data set were analyzed so far; the data were used only to measure the stellar dynamical mass through the CO bands in \citet{Onken14}. The exposure time of the observations was 120\,s, with a maximum flux variation of 4 per cent between the selected data cubes. We removed the telluric bands and calibrated the flux using the A0V star HD\,98152, with $H$-band magnitude of 8.89.
    
\begin{figure*}
    \resizebox{\hsize}{!}{\includegraphics{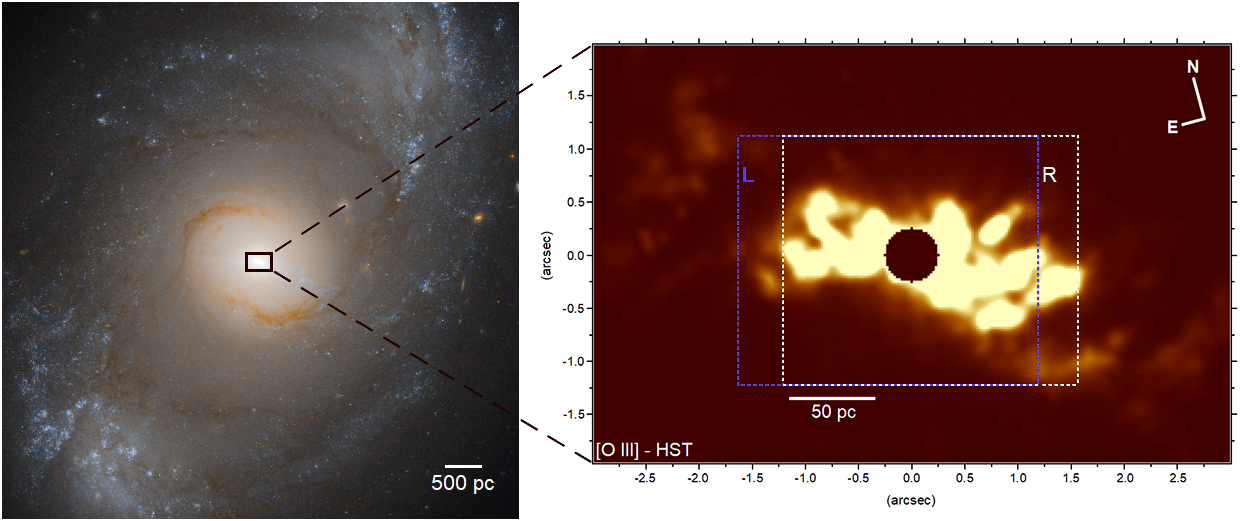}}
    \caption{HST filter composition of NGC\,4151 (credits to ESA/NASA/Judy Schmidt) and a zoom on the nuclear region showing the [O\,{\sc iii}] emission \citep{Hutchings99} with the the spatial arrangement of the NIFS observations (contours). The dashed blue and white rectangles depict the furthest left and right dithering, respectively, and their interception represent the combination of all data cubes.}
     \label{fig:fov}
    \end{figure*}

    The arrangement of the nine observations (three data sets with different dithering in the $x$-axis) is shown in Fig.~\ref{fig:fov} (right panel) and it is organized as follows: the left (\textbf{L}) and right (\textbf{R}) areas (coloured dashed rectangles) represent the median of 3 data cubes each (shifted by -0.42 and 0.38 arcsec from the centre, respectively); the central median data cube, despite being wider than the area intersected by \textbf{L} and \textbf{R}, has been cut to fit into the intersection of the two. The reason is to avoid two things: more borders and a ``messy'' image transition.   
    Therefore, the intersection of both areas (\textbf{L} and \textbf{R}) is a combination of all nine data cubes (through an average of 3 median data cubes) and the borders are just a median of 3 data cubes each. Although there is no dithering in the $y$-axis, we have cut a total of 15 spaxels in this dimension because of instrumental imperfections at the borders.
    
    The combination of data cubes significantly improves the signal-to-noise ratio (S/N) and the final FoV (3.17 $\times$ 2.35 arcsec) is a balance between data quality and presenting the main NLR structures.
    Since our focus is to highlight the weak structures of the emission lines in the NLR of NGC\,4151, we avoid them to be overshadowed by the central emission (the brightest region in the FoV) by applying a nuclear mask with a radius of 0.20 or 0.25 arcsec (depending on the image) centred in the galactic bulge intensity peak.
    
    The $K$-band data set has an average seeing of 0.85 arcsec (3 times worse than in the $H$-band) and a shorter individual exposure time of 90\,s. The standard star used for the data calibration was HD\,67201, with spectral type of A0 and $H$-band magnitude of 7.69. The final data cube has 8 combined observations, resulting in a FoV of 2,90 $\times$ 2.35 arcsec$^{2}$. The aim to present the $K$-band is, in addition to compare the data treatment with the $H$-band, to produce more accurate kinematic maps among the H$_{2}$ and H\,{\sc i} lines (which have more intense transitions than in the $H$-band).

    Finally, the K$_{long}$-band data set aims uniquely to present a brief analysis of the [Si\,{\sc vii}] coronal line (which was not analyzed so far and will be better explored by May D., et al., in preparation). Given its high ionization potential (IP), of 205\,eV, the [Si\,{\sc vii}] emission represents a distinct gas phase when compared to other studied emission lines, providing a more complete view of the NLR in NGC\,4151 (as shown for NGC\,1068 in M\&S17). Here, we limit the coronal emission analysis just for the comparison between its morphology with other emission lines. These observations were calibrated with the standard star HIP\,61471 (spectral type A0V and $H$-band magnitude of 7.34) and present the highest average seeing among our data (of 1.02 arcsec), being later combined in a data cube of 9 individual exposures of 90\,s each.

   \subsection{Data treatment}
   \label{sec:dt}
   
   A complete description of the data treatment for the specific case of NIFS data is given in the work of \citet{Menezes14}, which uses as a science case the $K$-band data of the galaxy NGC\,4151 itself.
   The following procedures were applied to all data sets: empirical correction of the differential atmospheric refraction (with a maximum centroid displacement of only 0.05 arcsec), spatial re-sampling with quadratic interpolation to 0.021 arcsec (half of the smaller detector pixel scale) and the Butterworth spatial filtering (with cut-off frequency \textsl{f}=0.35 and 0.40 in the $x$ and $y$ directions, respectively), to remove high spatial frequencies without affecting the point spread function (PSF).    
   
\begin{figure}
    \resizebox{0.85\hsize}{!}{\includegraphics{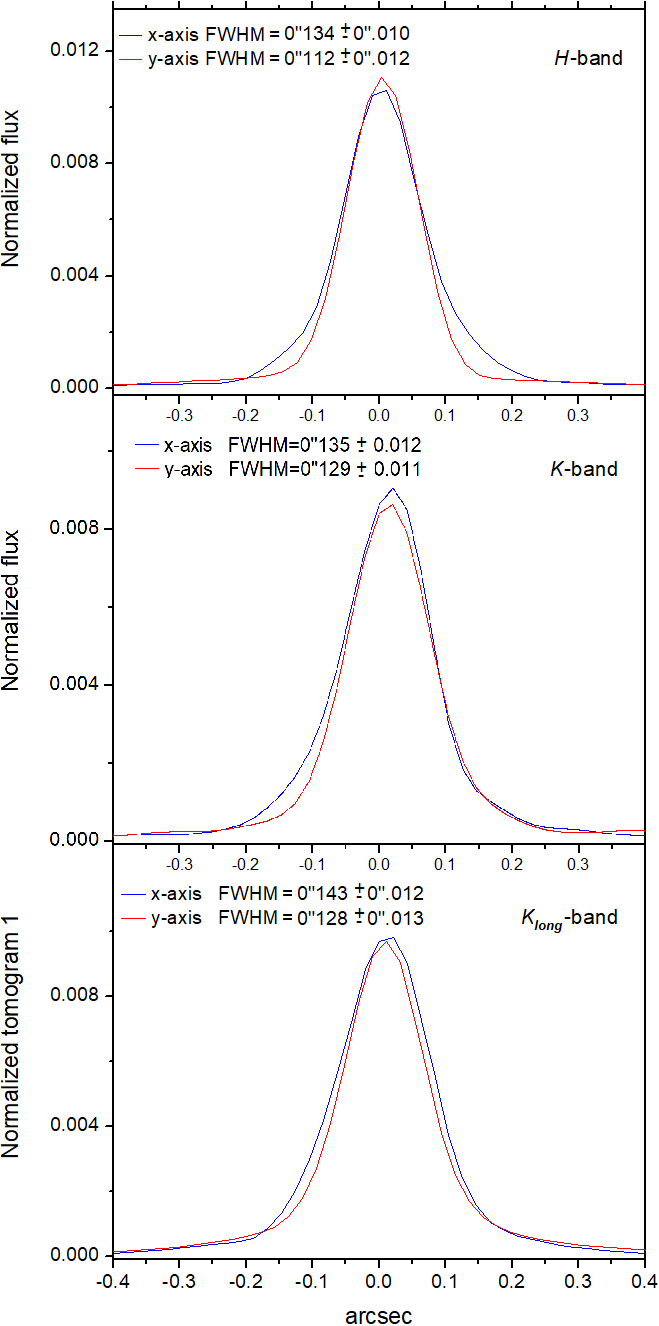}}
    \caption{PSF profile of the standard star for the $H$-band (top), of the broad Br$\gamma$ component for the $K$-band and of the tomogram 1 for the $K_{long}$-band (bottom).}
     \label{fig:psf}
    \end{figure}

\begin{figure*}
    \resizebox{\hsize}{!}{\includegraphics{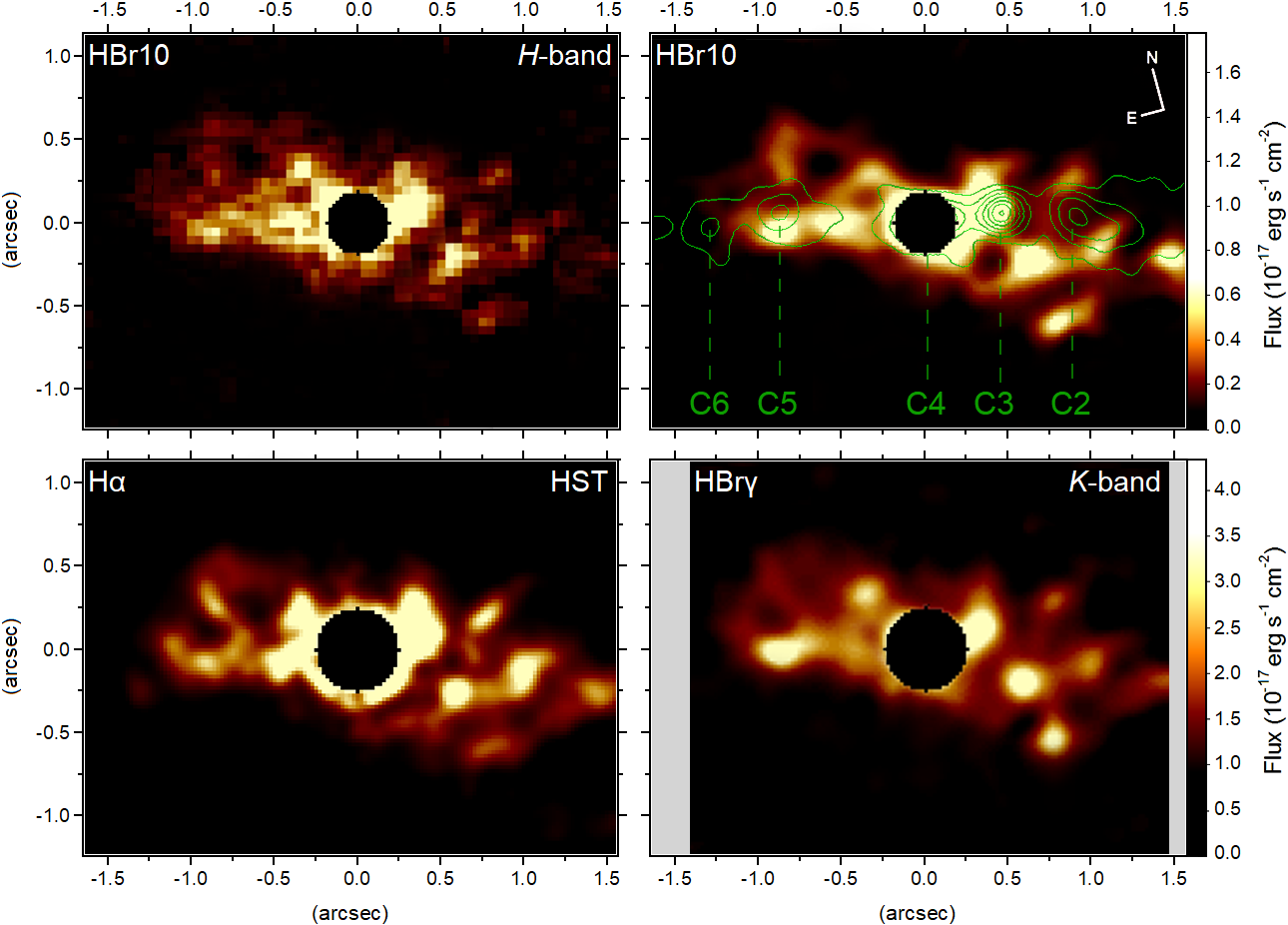}}
    \caption{Top-left panel: image of the $H$-band Br10 $\lambda$17367 \AA~emission line from 9 combined data cubes before the data treatment; top-right panel: the Br10 image after the data treatment overlaid with the contours of the 1.5\,GHz e-MERLIN radio emission; bottom-left panel: the H$\alpha$ image from the HST; bottom-right panel: the $K$-band Br$\gamma$ $\lambda$21661 \AA~emission after the data treatment.}
     \label{fig:hbr}
    \end{figure*}

   Once a PSF has been choose, the last step is the application of the Richardson-Lucy PSF deconvolution method \citep{Richardson72,Lucy74}.
   For the $H$-band, we could not find a reliable PSF from the data, then we have opted to use a PSF extracted from the standard star, similar to what was performed in \citet{DMay16}. The PSF profiles are shown in Fig.~\ref{fig:psf} (upper panel), with a FWHM slightly broader in the $x$-axis due to the original pixel scale in this direction. After 6 iterations, the final result of the data treatment is shown in Fig.~\ref{fig:hbr} (upper panels), where we compare the Br10 $\lambda$17367 \AA~emission line image before and after the treatment.
   
   For the case of the $K$-band, in turn, we were able to use the best strategy to apply deconvolution, which is the extraction of a PSF from the own data. For this purpose, the blue and red wings of the detected broad component in the Br$\gamma$ emission were selected and added together in one single image, composing the PSF (with spatial profiles shown in Fig.~\ref{fig:psf}, middle panel). In this case, 10 iteractions were applied and the final image is shown in Fig.~\ref{fig:hbr} (bottom right panel).
   
    In Fig.~\ref{fig:hbr} we also compare the impact of the data treatment between the $H$- and $K$-bands and with the H$\alpha$ image from the \textsc{HST} (obtained with the WFPC2 camera and the F658N filter). At first sight, all the images are notably similar between each other, with all the main structures seen in the \textsc{HST} image also visible in the NIFS images.
    At this point it is clear that, allied to the data treatment, the excellent seeing observations in the $H$-band led to a meaningful improvement of the IFU data cubes. One may note, for instance, two symmetric cavities close to the nucleus, at a PA$\sim$45\textdegree. Despite being better defined in Br10, they are still compatible with the H$\alpha$ emission from the \textsc{HST}.

    The spatial resolution reached in the $H$-band after the PSF deconvolution process can be estimated from a nuclear [Fe\,{\sc ii}] transient emission (Appendix~\ref{sec:tr}), which was not used as PSF before because it has a FWHM slightly above the PSF of the standard star. For this reason we cannot guarantee that this transient emission is, indeed, unresolved, but it is the closest point-like emission we can use to quantify the deconvolution impact. The resulting FWHM of the spatial profile from the transient emission (using the PSF of the standard star) is 0.128$\pm$0.009 arcsec, which is 8.3\,pc in the galaxy. This result represents an improvement of about 10 per cent when compared to the FWHM before the deconvolution (of 0.138$\pm$0.005).
    
    For the the $K$-band we measure a value of 0.102$\pm$0.008 arcsec (or 6.6\,pc) for the FWHM after the deconvolution. The improvement of the PSF in the $K$-band, where a point-like emission is found (i.e., the broad Br$\gamma$ emission) is of 35 per cent. Nevertheless, the final FWHM in the $K$-band is only slightly better than in the $H$-band, which has a much better average seeing. We should keep in mind that the PSF estimate is less reliable in the $H$-band, where the ionized H emission still seems to present sharper structures than in $K$-band (Fig.~\ref{fig:hbr}).
    Such detected structures have, in fact, a spatial resolution almost as good as the the diffraction limit of the Hubble Telescope for the H$\alpha$ wavelength, of 0.07 arcsec.
    It is worth noticing that the difference between the current Br$\gamma$ image and the one published by \citet{Menezes14} (using the same data) is attributed to the use of a slightly distinct PSF in the deconvolution method.

    Finally, our interest in the $K_{long}$-band is to show only the [Si\,{\sc vii}] morphology. Here, we selected a PSF following the same method described in M\&S17, where the Principal Component Analysis (PCA, \citealt{Steiner09}) is used to isolate the point-like hot dust emission. The corresponding spatial profiles of the PSF are shown in Fig.~\ref{fig:psf} (bottom panel).

    \section{Results}
    \label{sec:results}
    
    The main motivation of this work is to check the presence of an ``hourglass'' wall structure in NGC\,4151, as seen in the low-velocity [Fe\,{\sc ii}] emission of NGC\,1068, as well as if the high-velocity emission fills in the hourglass volume (M\&S17).
    The combination of excellent seeing observations with our routine of data treatment has revealed a new scenario for the later galaxy, which led us to adopt the same strategy in the case of NGC\,4151.
    In Fig.~\ref{fig:hbr}, we show, for instance, that the final image of Br10 reaches a resolution comparable to the H$\alpha$ image from the \textsc{HST} with the WFPC2 camera. Therefore, inspired by the results found in NGC\,1068, we start this section looking for a possible dichotomy between the low- and high-velocity [Fe\,{\sc ii}] emission.

    \subsection{The [Fe\,{\sc ii}] emission}
    \label{sec:FeII}
    
    \subsubsection{The low- and high-velocity NLR}
    \label{sec:felh}

\begin{figure*}
    \resizebox{\hsize}{!}{\includegraphics{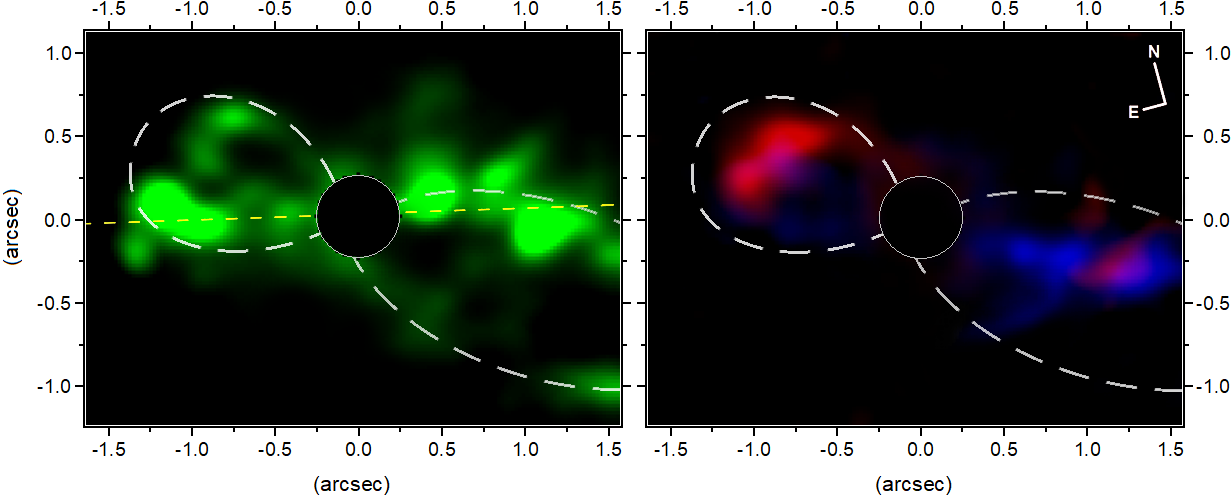}}
    \caption{Left panel: the low-velocity [Fe\,{\sc ii}] emission with the white contours delimiting the walls of an ``hourglass'' structure and the yellow dashed line the PA of the jet; right panel: RB for the high-velocity emission (see velocity ranges in Table~\ref{table:vel}).}
     \label{fig:Felh}
    \end{figure*}

    To check any dichotomy in the ionization cones of NGC\,4151, with respect to their velocity regimes, we first selected the low-velocity [Fe\,{\sc ii}] by summing together the three central velocity frames, shown in Fig.~\ref{fig:Felh} (left panel). This image clearly shows an hourglass-shaped structure with its walls roughly depicted by the dashed contours. (The NE end of the contours does not properly represents the low-velocity emission there, but the overall shape still is quite accurate to follow our argumentation through this work). The velocities were chosen to show the hourglass geometry and are close to the systemic velocity of the galaxy (-92<v<73~\kms; Table~\ref{table:vel}). This regime is $\sim3$ times lower than the low-velocity channels selected for NGC\,1068 but, there, all the velocity regimes are systematically larger. This is why we also choose different velocity ranges for the other emission lines, here.
    It is worth noticing that, although one can not confirm if the SW side of the hourglass is closed in on itself (because it goes beyond our FoV), it is clearly more extended than the NE side. We estimate a PA of 58\textdegree$\pm3$\textdegree for the low-velocity structure, which is in agreement with the orientation for the ionization cones in the literature (measured as 60\textdegree~by \citealt{Das06} and 56\textdegree$\pm6$\textdegree~by \citealt{Wang11a}).

    The high-velocity [Fe\,{\sc ii}] emission (see the velocity ranges in Table~\ref{table:vel}) is shown in Fig.~\ref{fig:Felh} (right panel) and one can see that, in the SW cone, the high-velocity is entirely within the hourglass structure depicted by the low-velocity emission (as represented by the dashed contours). In the NE cone, although it seems to trace more or less the low-velocity gas, the high-velocity gas is more receded from the side walls of the hourglass contours. Even with some overlapping, the high-velocity emission extends towards the inside of the NE cone, covering the inner walls of the low-velocity emission. The redshifted emission, for instance, is clearly thicker than the hourglass walls.
    
    There are more differences in the high-velocity emission distribution among both sides of the hourglass, like the SW structures being more distant from the hourglass walls and the presence of a curious arc-shape structure orthogonal to the hourglass axis (further discussed in Sect.~\ref{sec:arc}). On the other hand, at the NE the emission looks more filamentary. Such an asymmetry suggests distinct gas dynamics between both sides of the hourglass.
    Finally, despite of this dichotomy, one may find redshifted and blueshifted velocities on both sides of the hourglass' interiors, but with a preference for the NE side to show more redshifted emission and the opposite case in the SW side.
    
    The PA=58\textdegree~for the high-velocity emission is estimated to be the same as for the low-velocity.
    Curiously, both velocity regimes show some emission outside the NW borders of the hourglass, which is much more evident than for NGC\,1068.
    
\begin{figure*}
    \resizebox{\hsize}{!}{\includegraphics{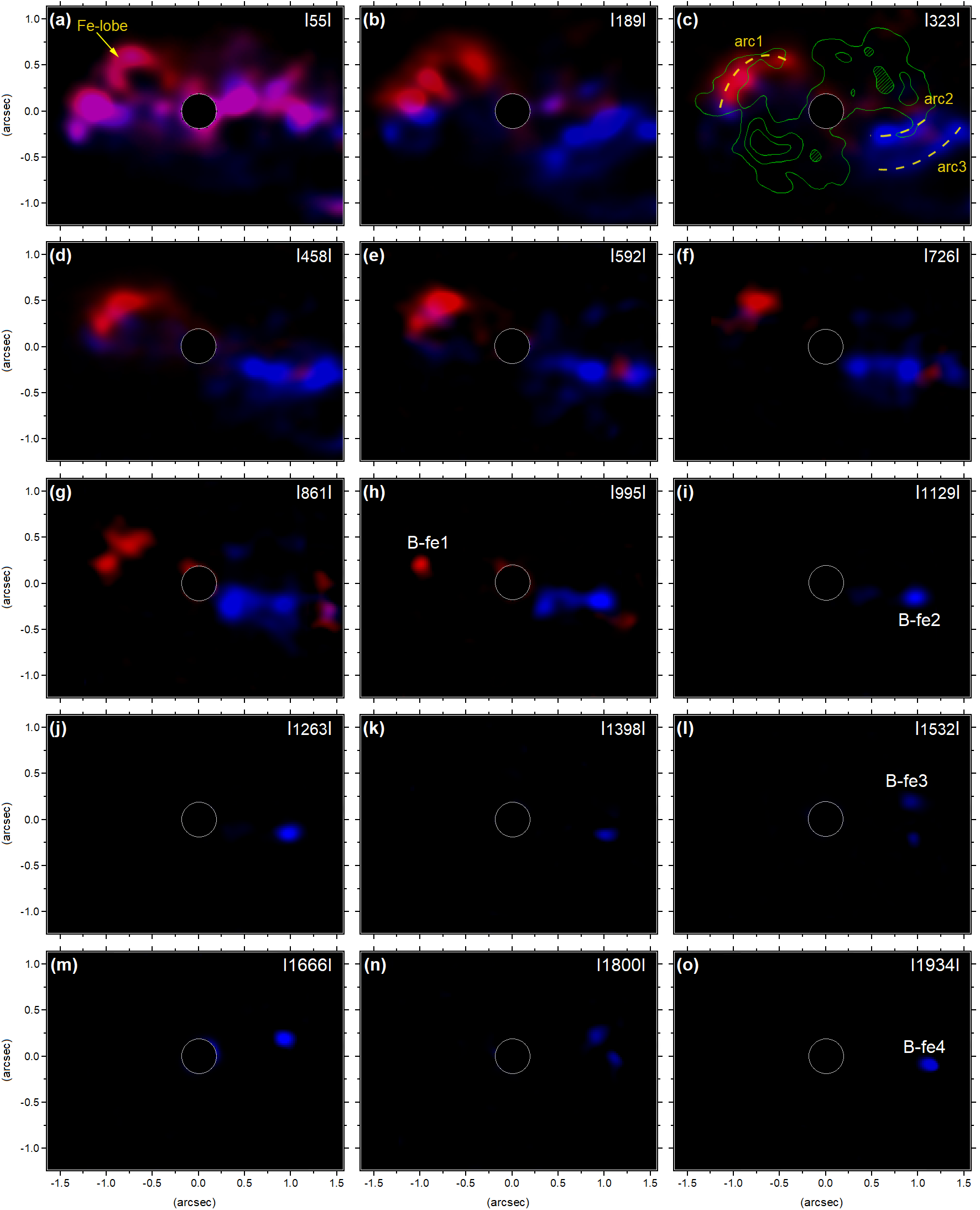}}
    \caption{BRV maps (blueshift and redshift velocity maps) for the [Fe\,{\sc ii}] $\lambda$16440 \AA~emission line. The velocity ranges from -2000 km s$^{-1}\lesssim v\lesssim$1000~km s$^{-1}$, with steps of $\sim$134 km s$^{-1}$. The green contours in panel \textbf{c} denote the molecular gas and the additional identifications are discussed in Sect.~\ref{sec:discussion}.}
     \label{fig:Febrvmaps}
    \end{figure*}

    The complex structure of [Fe\,{\sc ii}] may be better disentangled in the form of BRV maps (blueshifted and redshifted velocities), as shown by Fig.~\ref{fig:Febrvmaps}). These maps are obtained by summing together the velocity channels in a range ($\sim$134 km s$^{-1}$) around a certain velocity, with the same absolute velocities in blueshift and redshift at the same panels.
    Such maps are useful to highlight possible symmetries in the gas structures and kinematics.  
    
    We indicate in panel \textbf{a} of Fig.~\ref{fig:Febrvmaps} a lobe shape structure in the NE cone called ``Fe-lobe'', which is inside the NE side of the hourglass. This structure will be studied when presenting the H$_{2}$ data in Sect.~\ref{sec:h2kin}. 
    The BRV maps also show that, increasing in absolute velocities, the panels tend to reveal more fragmented structures, going from the low-velocity to the high-velocity regime. 
    Panels \textbf{b} and \textbf{c} show different structures among the two velocity regimes, drawing attention to the ones called ``arc1'' (in redshift at the NE cone) and ``arc2'' and ``arc3'' (two consecutive arcs in blueshift in the SW cone). 
    This means that the [Fe\,{\sc ii}] structures seen in the velocity range between the two regimes (which still represents low velocities when compared to the range selected in NGC\,1068) are not related to the hourglass walls nor to the fragmented structures inside these walls. After presenting the H$_{2}$ emission in Sect.~\ref{sec:h2}, we discuss that these arcs are, in fact, related to molecular ``barriers'' and are physically distinct from the low-velocity [Fe\,{\sc ii}].
    This smooth transition between the molecular and ionized Fe emissions is something new in NGC\,4151.

    One may see that the [Fe\,{\sc ii}] emission reaches velocities even higher (up to $\sim$2000 km s$^{-1}$) than those found by \citet{Hutchings99} for the [O\,{\sc iii}] emission (regions A, B, C and D in their Fig.4). For a radius lower than $\sim$1.5 arcsec (roughly our FoV), these regions are consistent with the locations of the structures with the highest velocities found for the [Fe\,{\sc ii}] emission. This finding is the first confirmation that the high-velocity clouds in the optical have a NIR counterpart.  
   
    Based on the faint SW blueshifted emission seen in panels \textbf{d} and \textbf{e} of Fig.~\ref{fig:Febrvmaps} (roughly at the cones' walls), a noticeable difference between the two cones may be noted, with the NE one presenting a more compact geometry and the SW part a more ``spread'' distribution for the gas clouds. These features are only possible to be inferred through the BRV maps (or channel maps) because all the velocity components are mixed together in velocity-collapsed images, where the faint structures are dominated by the strongest ones and become less evident or not visible at all. The same behaviour is more evident in the ionization cones of NGC\,1068, as can be seen in the image for the high-velocity emission (right panel of Fig.5 in M\&S17). There, where we reach a better spatial resolution (FWHM$\lesssim$0.1 arcsec), we tend to see more compact clouds, which lead to a better morphology characterization.
    In NGC\,1068 this geometrical difference between the cones is linked to the molecular gas layout surrounding the nucleus, a hypothesis that will be explored in Sect.~\ref{sec:bullets}.
    
    \begin{table*}
    \begin{center}
    \caption[vel]
    {Velocity range from the RGB composition of all emission lines presented in this paper, as shown in Figs.~\ref{fig:Felh}, \ref{fig:hbrk}, \ref{fig:h2} (right panel) and  \ref{fig:SiVII}.}
    \begin{tabular}{ccccc}
    \hline \hline
    $\lambda_{vac}$~(\AA) & Line & v$_{blue}^{a}$ (km s$^{-1}$) & v$_{green}^{a}$ (km s$^{-1}$) & v$_{red}^{a}$ (km s$^{-1}$) \\ \hline
    16 440 & [Fe\,{\sc ii}] & -2024$<v<$-383 & -92$<v<$73 & 401$<v<$1276 \\
    17 367 & H\,{\sc i}\,Br10 & -669$<v<$-224 & -121$<v<$147 & 249$<v<$684 \\
    17 480 & H$_{2}$ & -343$<v<$-111 & -69$<v<$86 & 120$<v<$292 \\
    24 833 & [Si\,{\sc vii}] & -736$<v<$-98 & -55$<v<$38 & 86$<v<$326 \\
    \hline
    \end{tabular}
    \begin{minipage}{15cm}
      Notes:
      (a) The uncertainty in velocity is $\sim$10 km s$^{-1}$. The different velocity ranges among the lines were chosen to better show a particular gas morphology for each line, as discussed in the text.
    \end{minipage}
    \label{table:vel}
    \end{center}
    \end{table*}

    \begin{table}
    \begin{center}
    \caption[bullets]
    {Peak velocity, velocity dispersion, distance from the nucleus and position angle measured for the gas bullets shown in Figs.~\ref{fig:Febrvmaps}, ~\ref{fig:Hbrbrvmaps}, ~\ref{fig:H2brvmaps} and ~\ref{fig:SiVII}.}
    \begin{tabular}{cccccc}
    \hline \hline
    Bullet ID & v$^{a}$ & $\sigma^{b}$ & Distance$^{c}$ & PA$^{d}$ & PA$_{c}$ $^{e}$ \\
     & (km s$^{-1}$) & (km s$^{-1}$) & (pc) & (\textdegree)  & (\textdegree) \\
    \hline 
    \multicolumn{6}{c}{[Fe\,{\sc ii}] $\lambda$\,16440 \AA} \\ \hline
    B-fe1 & 948 & 144$\pm$21 & 67 & 64 & 6 \\
    B-fe2 & -1221 & 163$\pm$16 & -65 & -115 & -7 \\
    B-fe3 & -1583 & 214$\pm$19 & -62 & -93 & -29 \\
    B-fe4 & -1867 & 219$\pm$20 & -77 & -110 & -12 \\ \hline
    \multicolumn{6}{c}{Br10 $\lambda$\,17362 \AA} \\  \hline
    B-br1 & 145 & 101$\pm$9 & 28 & 30 & -28 \\
    B-br2 & 411 & 107$\pm$15 & 58 & 43 & -15 \\
    B-br3 & -356 & 57$\pm$17 & -42 & -130 & 8 \\
    B-br4 & -318 & 61$\pm$10 & -66 & -139 & 17 \\ \hline
    \multicolumn{6}{c}{H$_{2}$ $\lambda$\,17480 \AA} \\ \hline
    B-h2a & 189 & 93$\pm$21 & 47 & 8 & -50 \\
    B-h2b & -208 & 91$\pm$14 & -37 & -155 & 33 \\
    B-h2c & 194 & 84$\pm$16 & 69 & 70 & 12 \\
    B-h2d & -230 & 70$\pm$17 & -58 & -119 & -3 \\ \hline
    \multicolumn{6}{c}{[Si\,{\sc vii}] $\lambda$\,24833 \AA} \\ \hline
    B-si1 & 100 & 41$\pm$5 & 31 & 31 & -27 \\
    B-si2 & -365 & 79$\pm$10 & -40 & -135 & 13 \\
    B-si3 & -362 & 60$\pm$8 & -65 & -142 & -20 \\
    B-s14 & -664 & 56$\pm$6 & -36 & -57 & 65 \\
    
    \hline
    \end{tabular}
    \begin{minipage}{8cm}
      Notes:
			(a) The uncertainty in velocity is $\sim$10 km s$^{-1}$.
            (b) Corrected for instrumental broadening.
			(c) Distance offset of each bullet from the bulge centre; negative numbers mean a southwest distance with respect to a line orthogonal to the cone's major axis.
            (d) Bullets' position angle relative to the northeast and (e) relative to the cone's major axis, of 58\textdegree~(or, equivalently, -122\textdegree) in the SW cone.
    \end{minipage}
    \label{table:bullets}
    \end{center}
    \end{table}
	
	Although we have numbered several gas clouds in NGC\,1068, the same strategy is not possible here because of the lower fragmentation degree for the gas structures shown in the BRV maps. Nevertheless, a pair of two remarkable compact gas clouds may be seen in panels \textbf{h} and \textbf{i}, namely B-fe1 and B-fe2 bullets (see properties in Table~\ref{table:bullets}). Both bullets are highly symmetrical with respect to their nuclear distances and proximity with the radio knots \textbf{C2} and \textbf{C5}. Despite of this proximity, the bullets direction relative to the nucleus is slightly tilted compared to the jet.
	The bullet B-fe4 is also very close to the bullet B-fe2, and it has an absolute velocity surprisingly larger, with a difference of $\sim$900 km s$^{-1}$ (or $\sim2$ times) from the bullet B-fe2. Thereby, the bullets represent the highest detected velocities for the [Fe\,{\sc ii}] clouds. A similar pair of bullets is also found in NGC\,1068, with nearly the same distances and an absolute velocity difference of $\sim$400 km s$^{-1}$. But there, they are and not associated with the jet. Apparently, the acceleration mechanism for the bullets in NGC\,4151 is more dramatic than NGC\,1068 and their origin will be discussed in Sect.~\ref{sec:discussion}. The [Fe\,{\sc ii}] bullet B-fe3 is also located at the vicinity of the radio knot \textbf{C2} and also present a high velocity of $\sim$-1600 km s$^{-1}$. In fact, all the blueshifted bullets seem to surround the radio knot \textbf{C2}.

    \subsubsection{The integrated [Fe\,{\sc ii}] emission}
    \label{sec:FeII} 
    
\begin{figure*}
    \resizebox{0.85\hsize}{!}{\includegraphics{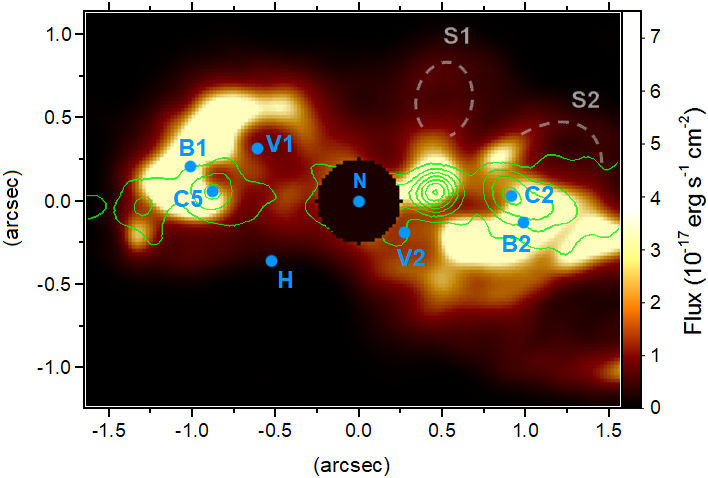}}
    \caption{Image of the [Fe\,{\sc ii}] $\lambda$16440 \AA~emission line overlaid with the e-MERLIN 1.5\,GHz radio emission (green contours). Contour levels are on a square root scale, from 67.3 to 1.0 mJy beam$^{-1}$. The blue letters identify locations corresponding to the spectra shown in Fig.~\ref{fig:spec} and the dashed lines indicate the structures \textbf{S1} and \textbf{S2}.}
     \label{fig:FeII}
    \end{figure*}

    The integral [Fe\,{\sc ii}] emission is shown in Fig.~\ref{fig:FeII} and depicts, also for all the velocity range, two well-defined ionization cones, as already seen in the previous section, with a PA=58\textdegree. From this angle, we can define an hypothetical orthogonal AGN obscuring torus, with a PA=148\textdegree. Differently from NGC\,1068, where the main [Fe\,{\sc ii}] peaks are distributed in a small region close to the nucleus (region \textbf{A} in Fig.3 of M\&S17), here they are spread all over the cones extension. We could highlight, however, two main arc-shaped structures located about 1 arcsec from the nucleus, suggestive of a symmetrical process. Equally curious are the structures \textbf{S1} and \textbf{S2} outside the ionization cones. They possibly represent gas cavities not directly associated with the central source. Considering the inclination of 45\textdegree~with respect to the line-of-sight \citep{Das05}, the real extension of the cones is estimated to be $\sim$280\,pc. This is $\sim$70\,pc less than the measured for NGC\,1068 (as already noted by \citealt{Cecil90} and \citealt{Das06}). 
    
    Overlaid to the [Fe\,{\sc ii}] image are the radio contours of the e-MERLIN  1.5\,GHz radio emission \citep{Williams17}. The radio knots are identified in Fig.~\ref{fig:hbr} (upper-right panel) and were named according to the literature (e.g. \citealt{Carral90}). The overlapping between the radio and all emission line images in this work were done matching the centroid of the nuclear radio component \textbf{C4} with a nuclear [Fe\,{\sc ii}] transient emission (in the $H$-band - Appendix~\ref{sec:tr}), which is coincident with the [Si\,{\sc vii}] kinematic centre (in the $K$-band - Sect.~\ref{sec:clb}). Both positions are indicative of the AGN location in the present data and they also coincide with the continuum emission peaks for the $H$ and $K$-bands.
    The existence of sub-peaks in component \textbf{C4} (observed with VLBA and named as \textbf{C4E} and \textbf{C4W} in \citealt{Ulvestad98}) are in the resolution limit of our data, therefore they do not affect our criteria to match the images. 
    
    Although there is no obvious association between the radio knots and the [Fe\,{\sc ii}] peaks it is worth noticing that the radio knots \textbf{C5}, \textbf{C3} and \textbf{C2} (Fig.~\ref{fig:FeII}) are located, at least in projection, very close to some [Fe\,{\sc ii}] structures, with the \textbf{C5} and \textbf{C2} knots right before the location where the cones seem to become very bright and shaped in arcs. The jet orientation also coincides with the direction where these arcs are more extended with respect to the nucleus (8 and 22\,pc more in the NE and SW cone, respectively). 
    Furthermore, Williams, at al. (submitted) argue that, relative to the radio observations published by \citet{Pedlar93}, the decrease in flux of component \textbf{C3} indicates a continuous adiabatic expansion, probably associated with the heating of the NLR gas. Such fact, together with the resolved ``banana shape'' emission in \textbf{C3} \citep{Mundell03}, is suggestive of a jet interacting with a dense cloud of gas close to the AGN.
    
\begin{figure*}
    \resizebox{\hsize}{!}{\includegraphics{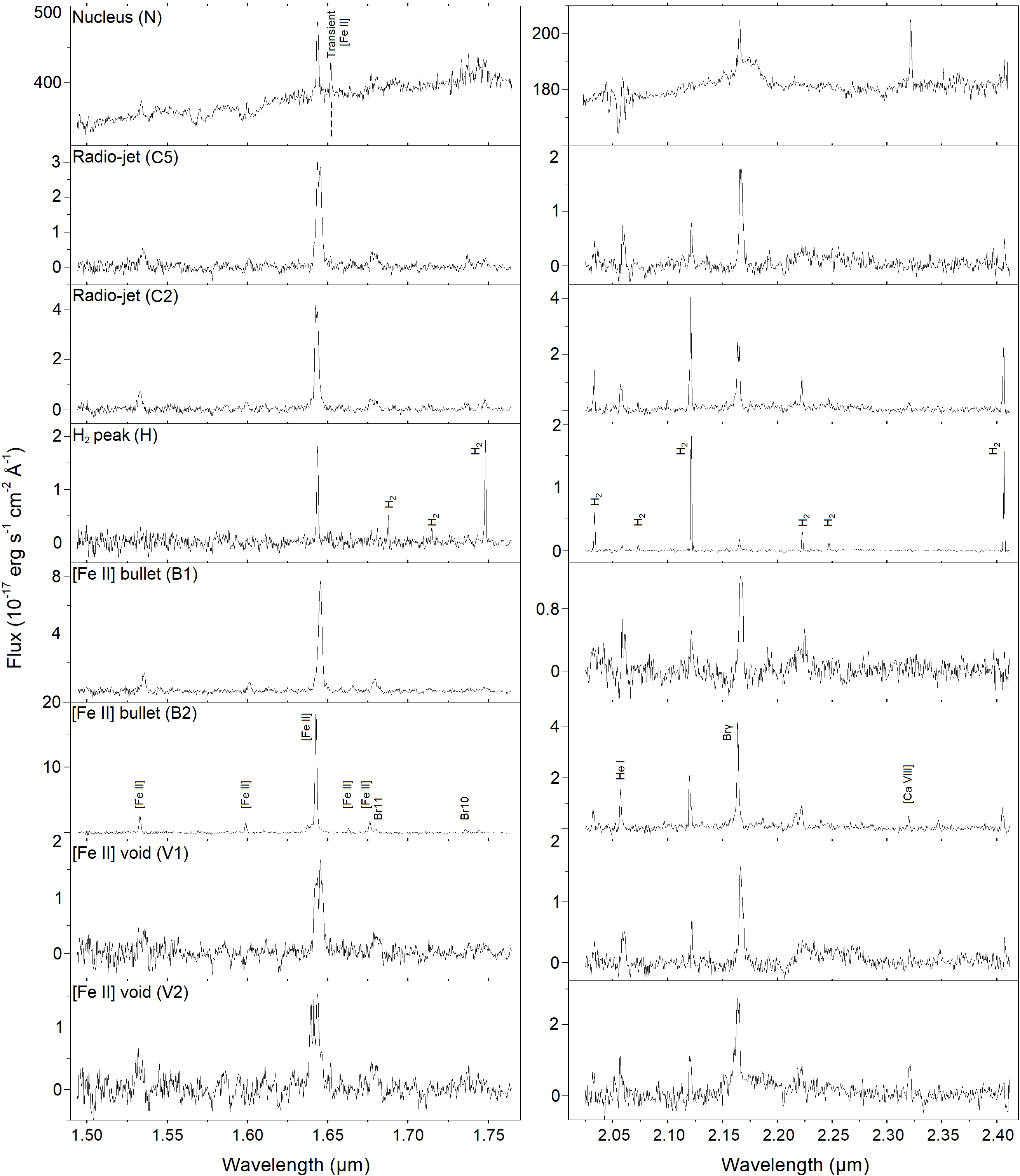}}
    \caption{Rest-frame spectra taken from eight regions marked in Fig.~\ref{fig:FeII}, with a circular aperture radius of 0.1 arcsec. For the nucleus (N) we used a larger aperture radius of 0.25 arcsec.}
     \label{fig:spec}
    \end{figure*}

    In Fig.~\ref{fig:spec} we compare the spectra extracted from some representative regions marked in Fig.~\ref{fig:FeII}, and in in Table~\ref{table:flux} we show the fluxes for the detected lines. It is important to keep in mind that the $H$-band observations used here were taken one year later than those of the $K$ and $H$-bands presented in \citet{Thaisa094151}. Such fact calls attention for two transient phenomena: 1) there is an emission line at $\sim \lambda$16520 \AA~only present in the spectrum shown here and 2) we do not see any sign of broad component in the profiles of the Br10 or Br11 emissions (as seen by \citealt{Thaisa094151}, one year before). The nature of the transient emission in (1) is explored in Appendix~\ref{sec:tr}. The broad line emission, in turn, is known to vary in scales of months \citep{Antonucci83,Penston84,Esser19} or completely fade away (\citealt{Boksenberg75} and this work).  
     
    The spectra taken from the radio knots \textbf{C5} and \textbf{C2} show double-peaked emission lines clearly seen in the [Fe\,{\sc ii}] and Br$\gamma$ profiles. The corresponding velocities of these peaks, for both lines, have an average difference of 246 km s$^{-1}$ (see Table~\ref{table:knots}), with a standard deviation of only 8 km s$^{-1}$. Such a result would hardly be a coincidence and it is suggestive of an expansion process driven by the jet in both sides of the cones.
    These pair of peaks are not associated with any bullet, but instead to the interface between the same knots and the arc-shaped [Fe\,{\sc ii}] structures. 
    
    \begin{table*}
    \begin{center}
    \caption[linefluxes]
    {Measured emission line fluxes for a circular aperture radius of 0.1 arcsec (except for the nucleus, of 0.25 arcsec) marked in Fig.~\ref{fig:FeII}, with the spectra shown in Fig.~\ref{fig:spec}. All values are in units of $10^{-15}$~erg s$^{-1}$~cm$^{-2}$. Blended lines were fitted by multiple Gaussian components and summed.}
    \begin{tabular}{ccccccccccc}
    \hline \hline
    $\lambda_{vac}$~(\AA) & Ion & \textbf{N} & \textbf{C5} & \textbf{C2} & \textbf{H} & \textbf{B1} & \textbf{B2} & \textbf{V1} & \textbf{V2} \\ \hline
    15 339 & [Fe\,{\sc ii}] & 4.46$\pm$0.43 & 0.18$\pm$0.02 & 0.20$\pm$0.02 & - & 0.34$\pm$0.03 & 0.40$\pm$0.05 & 0.05$\pm$0.03 & 0.85$\pm$0.12 \\
    15 999 & [Fe\,{\sc ii}] & - & - & 0.06$\pm$0.01 & - & 0.14$\pm$0.02 & 0.20$\pm$0.03 & - & - \\
    16 440 & [Fe\,{\sc ii}] & 19.21$\pm$0.73 & 2.09$\pm$0.07 & 1.38$\pm$0.09 & 0.23$\pm$0.04 & 0.04$\pm$0.01$^{a}$ & 2.70$\pm$0.03 & 0.85$\pm$0.08 & 0.56$\pm$0.07 \\
    16 642 & [Fe\,{\sc ii}] & - & - & - & - & 0.08$\pm$0.01 & 0.10$\pm$0.01 & - & - \\
    16 768 & [Fe\,{\sc ii}] & 3.96$\pm$0.86 & 0.12$\pm$0.03 & 0.020$\pm$0.004 & - & 0.18$\pm$0.02 & 0.28$\pm$0.04 & 0.14$\pm$0.04 & 0.26$\pm$0.9 \\
    16 811 & H\,{\sc i}\,Br11 & 4.15$\pm$0.98 & 0.11$\pm$0.03 & 0.02$\pm$0.01 & - & 0.09$\pm$0.03 & 0.12$\pm$0.02 & 0.11$\pm$0.03 & 0.22$\pm$0.6 \\
    16 877 & H$_{2}$ $1 - 0 S(9)$ & - & - & - & 0.039$\pm$0.005 & - & - & - & - \\
    17147 & H$_{2}$ $1 - 0 S(8)$ & - & - & - & 0.025$\pm$0.003 & - & - & - & - \\
    17 367 & H\,{\sc i}\,Br10 & 0.78$\pm$0.02 & 0.87$\pm$0.02 & 0.04$\pm$0.02 & 0.03$\pm$0.01 & 0.07$\pm$0.01 & 0.07$\pm$0.02 & 0.15$\pm$0.08 \\
    17 480 & H$_{2}$ $1 - 0 S(7)$ & 4.83$\pm$1.08 & 0.02$\pm$0.01 & 0.08$\pm$0.02 & 0.18$\pm$0.03 & 0.12$\pm$0.02$^{a}$ & 0.08$\pm$0.01 & 0.02$\pm$0.01 & 0.03$\pm$0.01 \\
    20 338 & H$_{2}$ $1 - 0 S(2)$ & - & 0.07$\pm$0.03 & 0.20$\pm$0.03 & 0.70$\pm$0.04 & - & 0.14$\pm$0.03$^{a}$ & 0.06$\pm$0.03 & 0.12$\pm$0.07 \\
    20 587 & He I & - & 0.18$\pm$0.05 & 0.24$\pm$0.02 & 0.13$\pm$0.04 & 0.16$\pm$0.05 & 0.31$\pm$0.03 & 0.17$\pm$0.05 & 0.28$\pm$0.06 \\
    20 735 & H$_{2}$ $2 - 1 S(3)$ & - & - & 0.03$\pm$0.01 & 0.12$\pm$0.04 & - & - & - & - \\
    21 218 & H$_{2}$ $1 - 0 S(1)$ & 4.23$\pm$0.78 & 0.21$\pm$0.08 & 0.69$\pm$0.01 & 2.39$\pm$0.06 & 0.12$\pm$0.04 & 0.44$\pm$0.08 & 0.11$\pm$0.03 & 0.24$\pm$0.07 \\
    21 661 & H\,{\sc i}\,Br$\gamma$ & 48.25$\pm$6.16 & 0.77$\pm$0.06 & 0.88$\pm$0.24 & 0.34$\pm$0.05 & 0.49$\pm$0.08 & 1.23$\pm$0.19 & 0.70$\pm$0.08 & 1.6$\pm$0.8 \\
    22 233 & H$_{2}$ $1 - 0 S(0)$ & - & 0.08$\pm$0.02 & 0.26$\pm$0.11 & 0.40$\pm$0.07 & 0.036$\pm$0.008 & 0.16$\pm$0.02$^{a}$ & - & 0.26$\pm$0.06 \\
    22 477 & H$_{2}$ $2 - 1 S(1)$ & - & 0.03$\pm$0.01 & 0.11$\pm$0.06 & 0.21$\pm$0.09 & - & 0.04$\pm$0.01$^{a}$ & - & - \\
    23 211 & [Ca\,{\sc viii}] & 44.78$\pm$2.49 & - & 0.08$\pm$0.03 & 0.08$\pm$0.05 & - & 0.09$\pm$0.03 & 0.02$\pm$0.01 & 0.22$\pm$0.08 \\
    24 066 & H$_{2}$ $1 - 0 Q(1)$ & - & 0.06$\pm$0.03 & 0.35$\pm$0.04 & 1.73$\pm$0.07 & - & 0.14$\pm$0.03$^{a}$ & 0.06$\pm$0.03 & - \\
    24 833 & [Si\,{\sc vii}] & 46.41$\pm$1.43 & 0.75$\pm$0.04 & 0.44$\pm$0.04 & 0.23$\pm$0.06 & 0.21$\pm$0.03 & 0.93$\pm$0.08$^{a}$ & 0.61$\pm$0.06 & 0.62$\pm$0.06 \\
    \hline
    \end{tabular}
    \begin{minipage}{15cm}
      Notes:
      (a) Flux measured only for the Gaussian component relative to the bullet emission.
    \end{minipage}
    \label{table:flux}
    \end{center}
    \end{table*}
    
     \begin{table*}
    \begin{center}
    \caption[knots]
    {Peak velocities for the double line components found for the [Fe\,{\sc ii}] and Br$\gamma$ emission lines in the spectra of Fig.~\ref{fig:spec} with regions \textbf{C5} and \textbf{C2} marked in Fig.~\ref{fig:FeII}.}
    \begin{tabular}{ccccccccc}
    \hline \hline 
    & \multicolumn{4}{c}{C5} & \multicolumn{4}{c}{C2} \\ \hline
    Line peak & [Fe\,{\sc ii}]\,1 & [Fe\,{\sc ii}]\,2 & Br$\gamma$\,1 & Br$\gamma$\,2 & [Fe\,{\sc ii}]\,1 & [Fe\,{\sc ii}]\,2 & Br$\gamma$\,1 & Br$\gamma$\,2 \\
     v (km s$^{-1}$) &
    -63 & 182 & -246 & -13 & -51 & 197 & -350 & -94 \\ \hline
    v$_{2}$-v$_{1}$ (km s$^{-1}$) & \multicolumn{2}{c}{245} & \multicolumn{2}{c}{233} & \multicolumn{2}{c}{248} & \multicolumn{2}{c}{256} \\ \hline
    \end{tabular}
    \label{table:knots}
    \end{center}
    \end{table*}
	
	It is interesting to note that, although regions \textbf{V1} and \textbf{ V2} in Fig.~\ref{fig:FeII} show low emission, they present the most complex line profiles of all FoV, with up to four visible components (Fig.~\ref{fig:spec}). According to the BRV maps (Fig.~\ref{fig:Febrvmaps}) such spectral features probably indicate uncoupled spatial components combined by projection effect. These regions lie along a PA of 15\textdegree~and are located right before the high-velocity emissions close to the jet borders.  
	
	There has been an attempt to probe the origin of the gas excitation through diagnostic diagrams in the NIR but, for the ionized gas, \citet{Ardila04} include the Pa$\beta$ line, which is not available in our data. On the other hand, \citet{Colina15} proposed to use only Br$\gamma$, with the [Fe\,{\sc ii}]/Br$\gamma$ vs. H$_{2}$/Br$\gamma$ line ratios. Basically, there is a trend to find an indication of ionization by the AGN with both line ratios becoming larger than unity. They have found a mean ratio of 4.56 and 0.78 for NGC\,4151, respectively (also with NIFS), and we obtain 4.8$\pm$0.2 and 0.7$\pm$0.1, which are compatible within the uncertainties. However, when calculated for the spatially resolved regions defined in Fig.~\ref{fig:FeII}, only region \textbf{H} lies unequivocally outside the star-forming locus of the diagram (Table~\ref{table:lineratio} and Fig.5 of \citealt{Colina15}). This is because the [Fe\,{\sc ii}] and H$_{2}$ emissions are more spread over the FoV and their sum exceeds the one of Br$\gamma$. In other words, if these ratios were calculated for a FoV more restricted to the Br$\gamma$ emission (such as the nucleus), the position for NGC\,4151 in the diagram would tend to move towards the star-forming locus. Therefore, there is a chance to misclassify the galactic nucleus because the measurements depend on where the Br$\gamma$ emission is enhanced by the outflow and from where (and with which aperture) the spectrum is extracted.

	\begin{table*}
    \begin{center} 
    \caption[lineratio]
    {Line ratios for some regions marked in Fig.~\ref{fig:FeII}, with the fluxes measured in Table~\ref{table:flux}.}
    \begin{tabular}{cccccc}
    \hline \hline
    Line ratio & $\frac{2-1 S(1)}{1-0 S(1)}$ & $\frac{1-0 S(0)}{1-0 S(1)}$ & $\frac{1-0 S(2)}{1-0 S(0)}$ & $\frac{1-0 S(1)}{Br\gamma}$ & $\frac{[Fe\,{\sc II}]}{Br\gamma}$ \\
    & & & & ($log$) & ($log$) \\ \hline
    \textbf{C5} & $0.14\pm0.07$ & $0.38\pm0.17$ & $0.88\pm0.44$ & $0.27\pm0.11$ & $2.71\pm0.23$ \\ 
    & & & & ($-0.57\pm0.18$) & ($0.43\pm0.04$) \\
    \textbf{C2} & $0.16\pm0.09$ & $0.38\pm0.16$ & $0.77\pm0.35$ & $0.78\pm0.21$ & $1.57\pm0.44$ \\ 
    & & & & ($-0.11\pm0.12$) & ($0.20\pm0.12$) \\
    \textbf{H} & $0.09\pm0.05$ & $0.17\pm0.03$ & $1.75\pm0.32$ & $7.03\pm1.05$ & $0.68\pm0.15$ \\ 
    & & & & ($0.85\pm0.06$) & ($-0.17\pm0.10$) \\
    \textbf{B2} & $0.1\pm0.3$ & $0.36\pm0.22$ & $0.88\pm0.17$ & $0.36\pm0.24$ & $2.20\pm0.34$ \\ 
    & & & & ($-0.44\pm0.29$) & ($0.34\pm0.07$) \\
    \textbf{N} & - & - & - & $0.09\pm0.02$ & $0.40\pm0.05$ \\ 
    & & & & ($-1.06\pm0.10$) & ($-0.40\pm0.06$) \\
    \hline 
    \end{tabular}
    \label{table:lineratio} 
    \end{center} 
    \end{table*}
    
    \subsection{The ionized hydrogen emission}
    \label{sec:hbr}
    
    In Fig.~\ref{fig:hbr} (upper-right panel) we show the Br10 $\lambda$17367 \AA~image after the data treatment, in the $H$-band, with the radio emission contours. One may see that the Br10 distribution is consistent with the ionization cones orientation defined by the [Fe\,{\sc ii}] emission. As a result, the PA for the ionized Hydrogen emission does not coincide with that of the radio jet. In fact, only a marginal relationship between the radio and Br10 emission may be seen, where the radio knots \textbf{C5} and \textbf{C3} seem to be close of two peaks of ionized gas.   
    
         \begin{figure}
  \resizebox{\hsize}{!}{\includegraphics{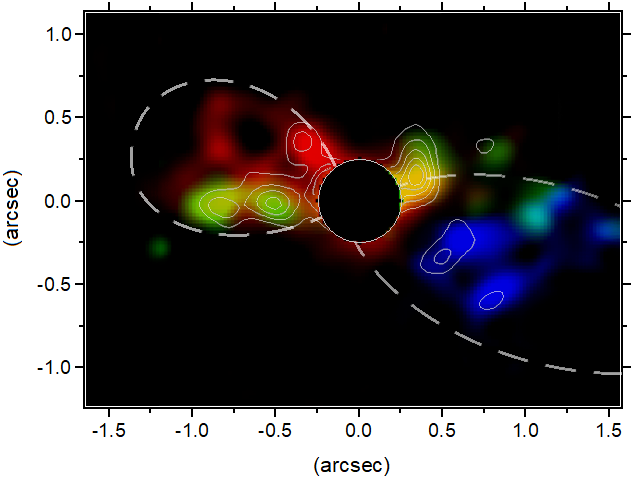}}
    \caption{RGB composition for the Br10 $\lambda$17362 \AA~emission line overlaid with the [Si\,{\sc vii}] contours (see velocity ranges in Table~\ref{table:vel}). The dashed contours delimits the hourglass structure seen in the low-velocity [Fe\,{\sc ii}] emission.}
     \label{fig:hbrk}
    \end{figure}
    
    According to the velocity RGB composition of Fig.~\ref{fig:hbrk}, the low-intensity emission reveals a more intricate structure of cavities fully contained inside the hourglass (except a fraction of the low-velocity emission in the SW cone), with almost no emission along its walls. Such features strengthen the fact that the low-velocity [Fe\,{\sc ii}] represents a partially ionized region and defines the cones' walls. 
    Now it becomes clear that the NE side of the cone exhibits nearly only redshifted emission and blueshift in the SW cone. Curiously, the low-velocity Br10 emission is preferably located along the jet PA.
    
\begin{figure*}
    \resizebox{\hsize}{!}{\includegraphics{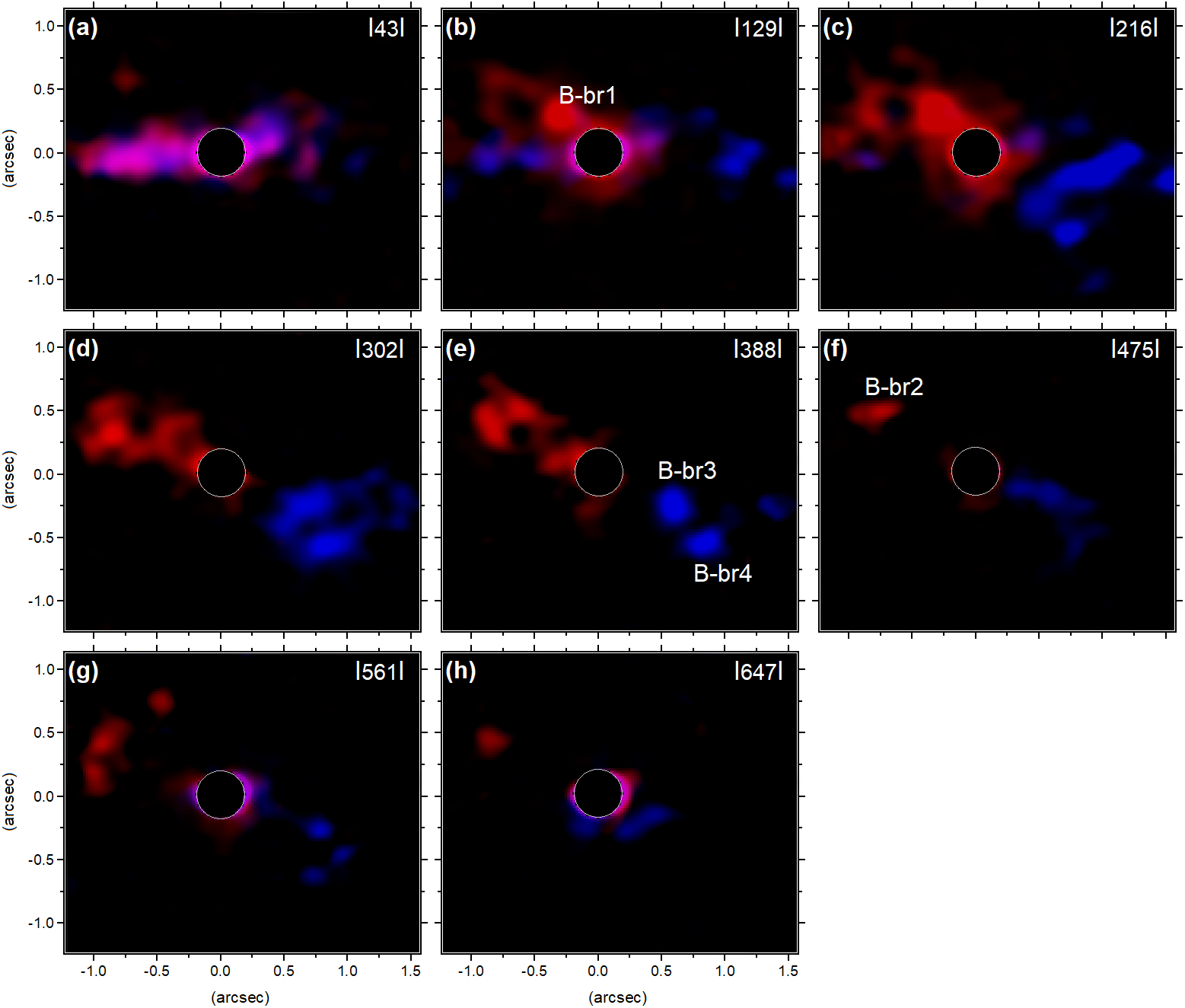}}
    \caption{BRV maps for the Br10 $\lambda$17362 \AA~emission line. The velocity ranges from -700 $\lesssim v\lesssim$700~km s$^{-1}$, with steps of $\sim$86 km s$^{-1}$.}
     \label{fig:Hbrbrvmaps}
    \end{figure*}

\begin{figure*}
    \resizebox{0.90\hsize}{!}{\includegraphics{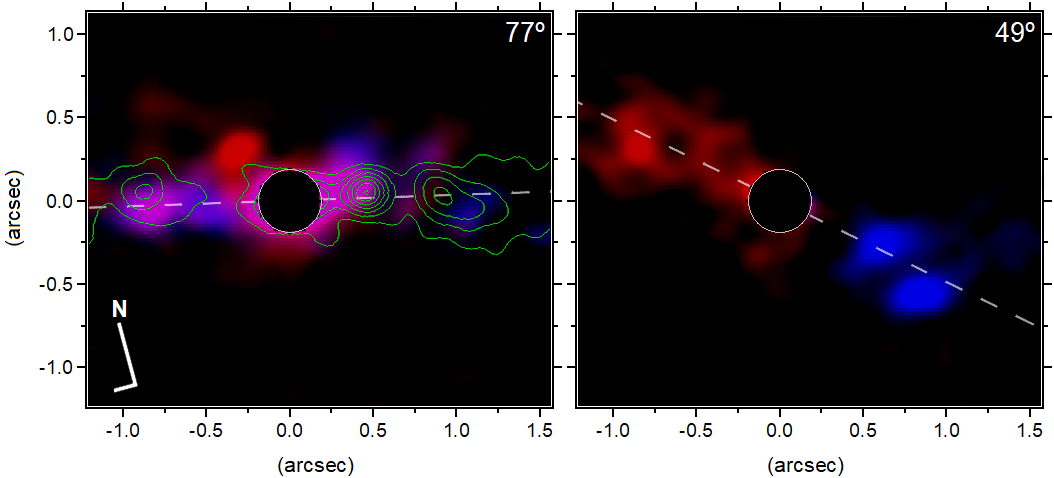}}
    \caption{Left panel: sum of panels \textbf{a} and \textbf{b} of the Br10 BRV maps from Fig.~\ref{fig:Hbrbrvmaps} ($\mid \overline{v}\mid=86$ km s$^{-1}$), with the radio emission (contours); right panel: sum of panels \textbf{d} and \textbf{e} from the same figure ($\mid \overline{v}\mid=345$ km s$^{-1}$). The angles denote the PA of the resulting gas distributions.}
     \label{fig:hbr_dual}
    \end{figure*}

    The complex Br10 distribution may be better analyzed through the BRV maps (Fig.~\ref{fig:Hbrbrvmaps}). At first, we note two velocity regimes: 1) structures close to the systemic velocity aligned with the radio jet, with velocities up to $\sim\mid v\mid = 200$ km s$^{-1}$ and 2) high-velocity structures, up to $\sim\mid v\mid = 700$ km s$^{-1}$, oriented in the same direction of that of the ionization cones (as already noted by \citealt{Thaisa104151}). Except for panel \textbf{c}, which seems to represent the transition between the two regimes, we define the low-velocity Br10, by adding panels \textbf{a} and \textbf{b} and the high-velocity with the sum of panels \textbf{d} and \textbf{e} in Fig.~\ref{fig:hbr_dual}. The resulting images have a PA of 77\textdegree~and 49\textdegree, respectively.
    Therefore, given the highly coincident PA for the jet and the low-velocity Br10, the presence of redshift and blueshift in both sides of the cones and the same velocity difference between the double peaks in \textbf{C5} and \textbf{C2} (see previous section), there are strong indications that the low-velocity regime represents the jet-gas interaction, possibly as the lateral expansion of the gas.
    
    Here, only the low-velocity Br10 is associated with the jet while in NGC\,1068 both velocity regimes are. The difference is that the jet and the cones have the same PA in NGC\,1068, calling attention to the fact that the jet might not be a requirement to accelerate the gas at high speeds, when present. Such fact increases even more the importance of an efficient momentum transfer mechanism to the NLR gas.
    
    From panel \textbf{c} of Fig.~\ref{fig:Hbrbrvmaps} onwards, the Br10 emission in blueshift and redshift starts to appear only in one side of the cones and present, along the panels, a good symmetry both in velocity and distance from the nucleus. The only exception is found in the last panel, where the highest redshifted velocity cloud remains distant from the nucleus, but the most blueshifted one becomes closer in the SW cone (such behaviour is also confirmed for the Br$\gamma$ in the $K$-band). 
    There are at least three gas bullets (compact gas cloud emission), identified as B-br1, B-br2 and B-br3 (see properties in Table~\ref{table:bullets}). The bullet B-br3 presents the highest velocity, in blueshift, with an intermediate nuclear distance from that of the bullets in blueshift, indicating a possible deceleration mechanism for the gas at 66\,pc from the nucleus in the SW cone. Although the kinematic modeling of \citet{Das05} has detected a deceleration only for a distance of 96\,pc from the nucleus, one may note a relevant fraction of clouds that divert from the expected behaviour exactly along the slit's position shown in \citet{Das05} that comprise the selected bullets.
    
    \subsection{The H$_{2}$~molecular emission}
    \label{sec:h2}

\begin{figure*}
    \resizebox{\hsize}{!}{\includegraphics{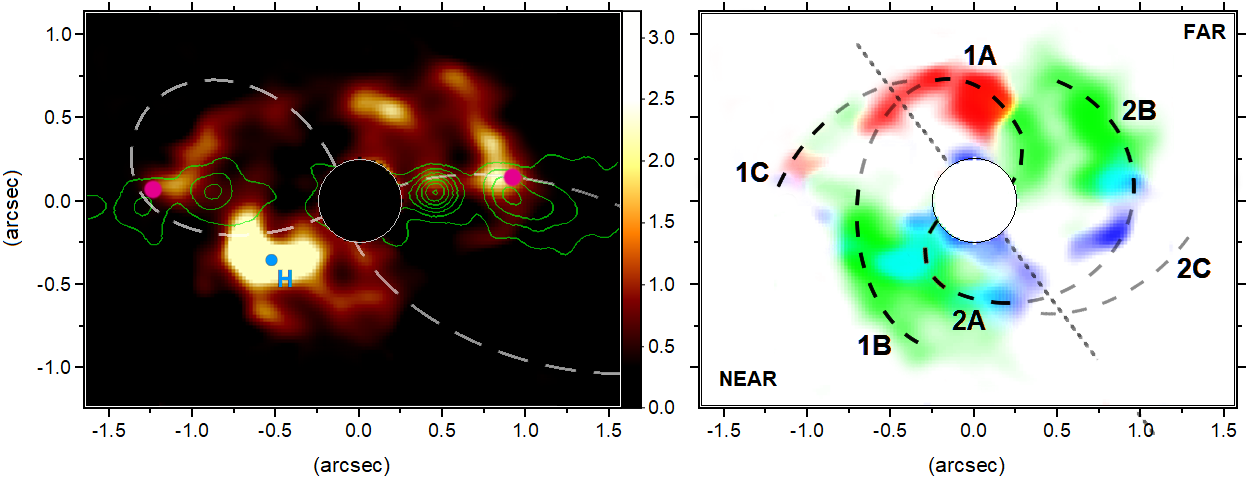}}
    \caption{Left panel: image of the H$_{2}$ $\lambda$17480 \AA~line ($H$-band) overlaid with the radio emission and the hourglass contours. The purple circles denote the [Ne\,{\sc ix}]/[O\,{\sc vii}] ratio peaks and the blue one the location of the extracted spectra shown in Fig.~\ref{fig:spec}. The flux units are in 10$^{-16}$ erg s$^{-1}$ cm$^{-2}$. Right panel: RGB composition for the same line (see velocity ranges in Table~\ref{table:vel}) with some identified arc-shaped structures (hereafter called ``segments''). The straight dashed line denotes the galaxy major axis, of 22\textdegree, showing where is the near and far sides of the galactic disc.}
     \label{fig:h2}
    \end{figure*}

    In Fig.~\ref{fig:h2} (left panel) we show the molecular gas distribution with the contours of the radio emission (green) and of the hourglass walls (white) seen in the low-velocity [Fe\,{\sc ii}] emission.

    Unlike the other emission lines, the H$_{2}$ has a distribution preferably located in the shadow of the ionization cones. The apparent lack of H$_{2}$ in the region of the cones, with such a well-defined limit between the inner and outer parts of the hourglass, may indicate that a major fraction of molecules exposed to the central source were already destroyed. 
    
    Although the hourglass roughly delimits the borders of the H$_{2}$ outside the contours, the molecular structure is visibly not perpendicular to the cones/hourglass axis, with an estimated PA of $\sim$121\textdegree ~(27\textdegree less than that expected for a torus, as inferred from the PA of the ionization cones). Curiously, the axis defined by the high-velocity Br10 (Fig.~\ref{fig:hbr_dual}, right panel) is also smaller than the hourglass PA (by $\sim$10\textdegree). A similar, but more noticeable, effect is seen in NGC\,1068. There, aside the [Fe\,{\sc ii}] emission - that defines the ionization cones' PA - all the ionized gas is clearly concentrated in a distinct PA from that of the cones and also coincides with a lack of CO emission \citep{Burillo14}.
    
    In NGC\,1068 this difference in PA is explained by a misalignment between the torus and the accretion disc but here, even if the misalignment between the jet and the cones is real (Sect.~\ref{sec:desalin}), the accretion disc polar axis would not be aligned correctly to explain these features. Despite some wiggles seen in the highest resolution radio images \citep{Ulvestad98}, there is no sign of a bent jet at sub-arcseconds scales in NGC\,4151 (what could place the accretion disc at another orientation). Therefore, nothing more can be said at this time about this apparent discrepancy between NGC\,4151 and NGC\,1068.
    
    It is interesting to note the apparent coincidence of the radio knots \textbf{C3} and \textbf{C2} (Fig.~\ref{fig:hbr}, upper-right panel) with the tips of the molecular arcs seen in the SW cone (Fig.~\ref{fig:h2}, left panel). In NGC\,1068 there are also two radio knots located exactly where one may find two molecular walls (Fig.18 of M\&S17). In the NE cone of NGC\,4151 the radio knots \textbf{C6} and \textbf{C5} seem to be in an advanced position with respect to the molecular arcs, which, in turn, are also more distant from the nucleus.
   
    The curious ``double shell'' structure of molecular arcs in the overall H$_{2}$ distribution draws a lot of attention to what might be its nature, in fact.
    It is interesting to note that its NS dimension matches very well with the H\,{\sc i} ring proposed by \citet{Mundell03} to explain the obscuration towards the radio sub-component \textbf{C4E} (eastern side of the nucleus - \textbf{C4W}).
    In Sects.~\ref{sec:desalin} and \ref{sec:bullets} we explore the possibility of the molecular gas be affected by the outflow and its real distribution within the galactic disc. 
    
    Following the analysis presented by \citet{Mouri94}, we may probe the molecular gas excitation through the H$_{2}$ ratios that were possible to be measured from our data (i.e., $1-0 S(2)$/$1-0 S(0)$, $2-1 S(1)$/$1-0 S(1)$ and $1-0 S(0)$/$1-0 S(1)$). However, only half of the selected regions in Fig.~\ref{fig:FeII} (\textbf{C5}, \textbf{C2}, \textbf{H} and \textbf{B2}) have the required H$_{2}$ emissions to calculate these ratios. Both diagrams in Figs.1 and 3 of \citet{Mouri94} are equivalent but, given the uncertainties, the ratios $2-1 S(1)$/$1-0 S(1)$ and $1-0 S(2)$/$1-0 S(0)$ of Table~\ref{table:lineratio} provides the most accurate constraints. The results comprise two types of behaviours: while all the values for regions \textbf{C5}, \textbf{C2} and \textbf{B2} lie very close to each other, with temperatures $<$1000\,K, region \textbf{H} (centred on the H$_{2}$ peak) lies precisely on the curve predicted by X-ray excitation where the temperature is 2000\,K. Region \textbf{H} is located at the external edge of the NE cone and may be heated by X-ray photons crossing the borders of the central dusty torus, which penetrate more deeply into the gas. The other regions, in turn, are consistent with thermal UV excitation plus an additional heating mechanism, such as shocks and/or X-ray excitation, again. This is somewhat expected, since they are exposed to the UV radiation from the central source and are located along the jet, close to the radio knots.
    
    Finally, we can estimate the total H$_{2}$ mass of the detected structure, including the masked region, from the equation derived in  \citet{Scoville82} and \citet{Riffel08}:
    
     \begin{equation}
    \begin{aligned}
    M_{H_{2}}&=\frac{2m_{p}F_{H_{2}\lambda 2.1218}4\pi D^{2}}{f_{\nu =1,J=3}A_{S(1)}h\nu}\nonumber \\
    &=5.0776\times 10^{13}~\left(\frac{F_{H_{2}\lambda 2.1218}}{erg s^{-1}cm^{-2}}\right)\left(\frac{D}{Mpc}\right)^{2}~\Msun
    \label{h2mass}
    \end{aligned}
    \end{equation}

    \noindent where $m_{p}$ is the proton mass, $F_{H_{2}\lambda 2.1218}$ is the line flux, $D$ is the adopted galaxy distance, of 13.3\,Mpc, and $f_{\nu =1,J=3}$ is the fraction of hot H$_{2}$ in the level $\nu=1$ and $J=3$ , with $M_{H_{2}}$ given in solar masses. The linear dependence of the H$_{2}$ emissivity on density assumes T=2000\,K and $n_{H_{2}}>10^{4.5}$ cm$^{-3}$. This implies a population fraction of $1.22\times10^{-2}$ with transition probability $A_{S(1)}=3.47\times10^{-7}$ s$^{-1}$. According to the measured line ratios, the only regions that seem to deviate from a thermal equilibrium, at T$\sim$2000\,K, are those close to the radio knots, which correspond for less than 5 per cent of the total flux.
    From a measured flux of $F_{H_{2}\lambda 2.1218}=4.78 \times 10^{-14} erg~s^{-1} cm^{-2}$ we obtain a total mass of $M_{H_{2}}=429 \Msun$. According to \citet{Fiore17} and \citet{Flu19}, the molecular outflow mass is usually found to be larger than the ionized one, and in Sect.~\ref{sec:bullets} we take this into account to explore the energetics involved for each outflow phase.
    
    \subsubsection{H$_{2}$ kinematics}
    \label{sec:h2kin}
    
    A first analysis of the molecular gas kinematics is shown in the velocity RGB composition of Fig.~\ref{fig:h2} (right panel), which follows the same trend of redshifted and blueshifted emissions in the NE and SW sides of the cones, respectively (see velocity ranges in Table~\ref{table:vel}). The curved dashed lines are an attempt to connect the molecular ``double shell'' in a symmetrical structure, with segments \textbf{1A} and \textbf{2A} connected to the nucleus. This particular choice will be better justified in Sect.~\ref{sec:discussion}. The faded contours segments (\textbf{1C} and \textbf{2C}) represent the locations where the H$_{2}$ emission is weak or nearly absent.
    The region with the most prominent emission (Fig.~\ref{fig:h2}, left panel), for instance, would be part of two distinct segments (\textbf{1B} and \textbf{2A}), possibly merged by effect of spatial resolution.
    
\begin{figure*}
    \resizebox{0.65\hsize}{!}{\includegraphics{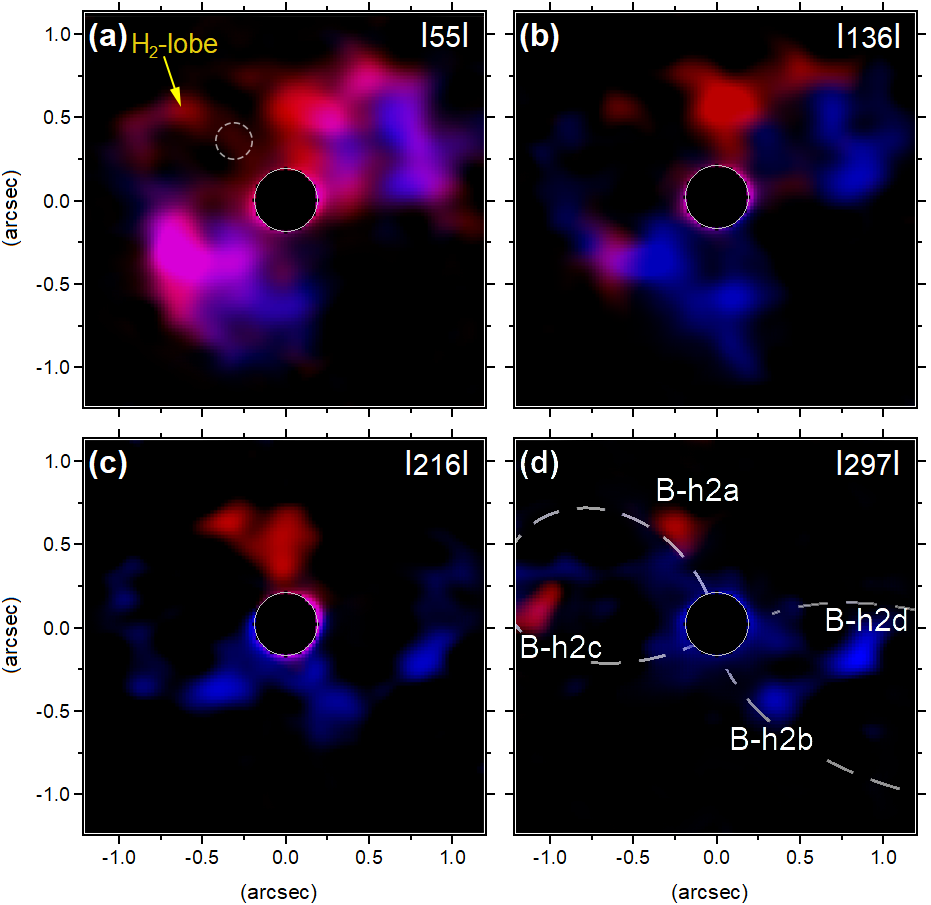}}
    \caption{BRV maps for the H$_{2}$~$\lambda$17480 \AA~emission line. The velocity ranges from -300 km s$^{-1}\lesssim v\lesssim$300~km s$^{-1}$, with steps of $\sim$80 km s$^{-1}$. The dashed circle in panel \textbf{a} denote the location of the [Si\,{\sc vii}] bullet B-si1 (Fig.~\ref{fig:SiVII}) and the contour in panel \textbf{d} the hourglass structure.}
     \label{fig:H2brvmaps}
    \end{figure*}

    Comparing both panels of Fig.~\ref{fig:h2} one may note that most of the emission concentrates close to the systemic velocity.
    In fact, the low-velocity H$_{2}$ emission mostly concentrates in the region outside the hourglass, almost perpendicular to it.
    To reveal fainter emissions, the BRV maps are shown in Fig.~\ref{fig:H2brvmaps} and already in panel \textbf{a} it is possible to see a distinct structure called ``H$_{2}$-lobe'' within the NE cone. One may ask if this structure is real, but, surprisingly, it matches both in position and velocity the Fe-lobe seen for the [Fe\,{\sc ii}] emission in \textbf{a} of Fig.~\ref{fig:Febrvmaps}, indicating that they are likely dynamically related. In panels \textbf{b} and \textbf{c} one may also observe cavity-like emissions on both sides of the cones.
    
\begin{figure}
    \resizebox{0.83\hsize}{!}{\includegraphics{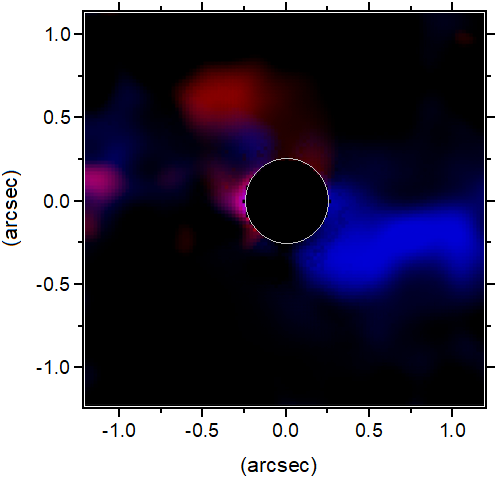}}
    \caption{Sum of the high-velocity channels ($|v|\sim$300 km s$^{-1}$) from the $K$-band H$_{2}$~$\lambda$20338 \AA~+ $\lambda$21218 \AA~+ $\lambda$22234 \AA~+ $\lambda$24085 \AA~emissions lines.}
     \label{fig:h2k}
    \end{figure}

    The most unexpected result, however, is that the high-velocity H$_{2}$ (panel \textbf{d} of Fig.~\ref{fig:H2brvmaps}) is almost totally seen in the region within the ionization cones/hourglass structure. Once more one may question the validity of such emission. For this reason, we added the same velocity channels of all the detected H$_{2}$ lines, in the $K$-band, in one single image. The resulting image is shown in Fig.~\ref{fig:h2k} and, despite the lower resolution, we detected an emission in the same regions of the structures seen in the $H$-band. Therefore, it is safe to claim that we detect, for the firs time in NGC\,4151, molecular gas exposed to the central source and that it is distinguished by its high-velocity emission.
    Such a finding is not a novelty for NGC\,1068, where compact H$_{2}$ clouds were found with velocities up to $|v|\sim$600 km s$^{-1}$ (\citealt{Burillo14}, M\&S17), twice as high as those detected in NGC\,4151.
    
\begin{figure}
    \resizebox{0.95\hsize}{!}{\includegraphics{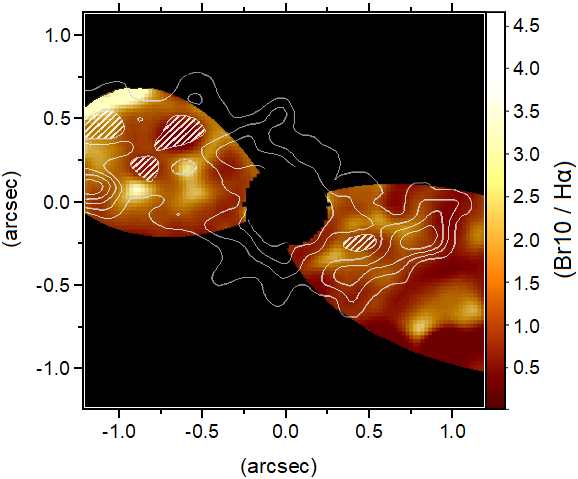}}
    \caption{Extinction map estimated from the $Br10/H\alpha$ ratio with the high-velocity H$_{2}$ emission (contours) from panel \textbf{d} of Fig.~\ref{fig:H2brvmaps}. Only the region inside the hourglass is shown.}
     \label{fig:dust}
    \end{figure}

    A second verification to strengthen the existence of the high-velocity molecular gas consists in the estimation of the extinction within the cones (through the Br10/H$\alpha$ ratio, which have comparable spatial resolutions - Sect.~\ref{sec:dt}), since the H$_{2}$ molecules still may be associated with dust in case of lower temperatures ($\lesssim1500$\,K - \citealt{Alexander83}) inside the ionization cones. The extinction map may be seen in Fig.~\ref{fig:dust}, with the contours of the high-velocity H$_{2}$ emission in panel \textbf{d} of Fig.~\ref{fig:h2k}. There are good indications that the absence of dust (darker regions) are also related to a lack of H$_{2}$ emission on both sides of the cones and there is at least one H$_{2}$ blob at $(x,y)=$(-0.5, 0.25 arcsec) associated with a higher extinction region. One possibility is that the high-velocity H$_{2}$ within the cones could be a projection effect, with a distribution outside the cones walls. According to the NLR geometry of \citet{Das05} and panel \textbf{d} of Fig.~\ref{fig:H2brvmaps}, this would necessarily imply redshifted bullets in the back wall of the NE cone and blueshifted ones in the front wall of the SW cone. Oddly enough, this distribution goes against the fact that these walls are much farther from the galactic disc in the proposed geometry (where no H$_{2}$ is expected) \citep{Das05}. We explore in Sects.~\ref{sec:arc} and ~\ref{sec:polardust} that this particular scenario is unlikely to be the case. The only reasonable explanation is that high density regions shield the dust from being destroyed by the AGN radiation.   
    
    We can discriminate four H$_{2}$ bullets nearly symmetrical with respect to the nucleus (see panel \textbf{d} of Fig.~\ref{fig:H2brvmaps} and properties in Table~\ref{table:bullets}), forming the counterparts (B-h2a - B-h2b) and (B-h2c - B-h2d). All of them have, within the estimated uncertainty, similar velocities. The redshifted bullets in the NE cone, however, are systematically farther and present a wider opening angle in the NLR. 
    As seen in Fig.~\ref{fig:FeII} (left panel), the NE cone is clearly less extended, what seems to be inconsistent with the location of the NE bullets. In NGC\,1068, in turn, the molecular ``barrier'' closest to the nucleus undergoes a process of fragmentation, while the most distant side expands without any disruption. Something similar might occur in NGC\,4151, with the closest bullets (SW cone) outflowing as a result of a past event of fragmentation and the farthest ones (NE bullets) still under the same process. In this way, the gas structure that still harbours the NE bullets had a longer period of expansion and, consequently, are more distant now. In other words, where the molecular structure disrupted first, the energy ``leaking'' prevented the gas clouds from being accelerated as efficiently as the opposite side. In Sect.~\ref{sec:bullets} we explore the origin of the bullets. 
    
    \subsection{Gas rotation}
    \label{sec:kin}
    
\begin{figure}
    \resizebox{\hsize}{!}{\includegraphics{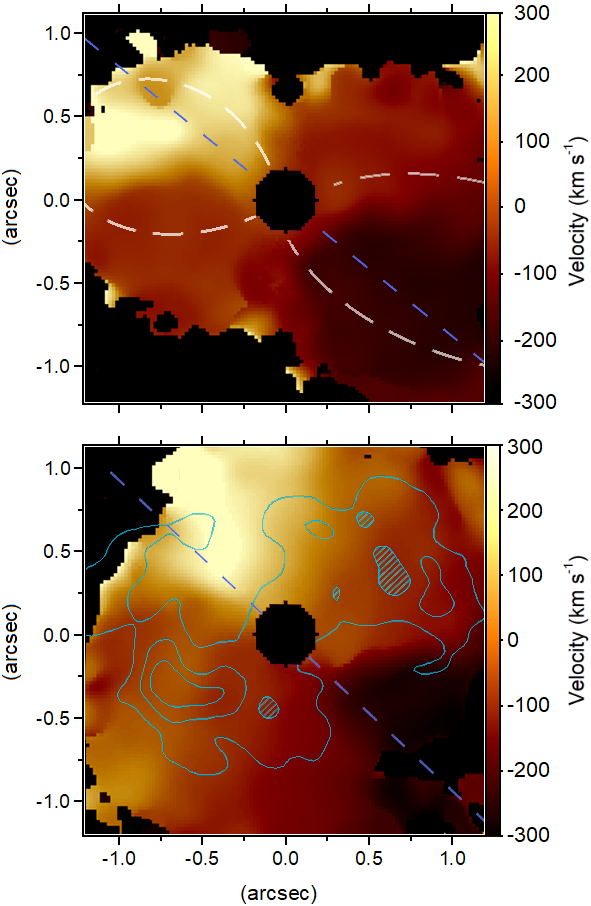}}
    \caption{Radial velocity maps for the Br$\gamma$ $\lambda$21661 \AA, with the hourglass contours (upper panel), and for the H$_{2}$ $\lambda$21218 \AA, with the contours of the integrated emission (bottom panel). The blue dashed lines denote the respective kinematic PAs.}
     \label{fig:rot}
    \end{figure}

    Despite of the complexity of the emission line profiles (Fig.~\ref{fig:spec}), we attempted to fit a Gaussian profile to the most intense line peak of the H\,Br$\gamma$ and H$_{2}$ $\lambda$21218 \AA~multiple emissions. Both fittings are performed in the $K$-band, where the transitions are more intense, with a higher S/N. The result is shown in Fig.~\ref{fig:rot}, with the measured PAs of 36\textdegree$\pm$2\textdegree~and 32\textdegree$\pm$5\textdegree, respectively.
    
    At first sight, the ionized gas does not seem to follow the orientation of the ionization cones (Fig.~\ref{fig:rot}, upper panel), but one may take into account that the kinematic map may be inaccurate in the region where we find the systemic velocity emission (see the double peaked lines in the spectra \textbf{C2} and \textbf{C5} in Fig.~\ref{fig:spec}). The peak related to lower velocities becomes more intense for a PA similar to that of the jet (77\textdegree), possibly resulting in a slightly smaller PA for the Br$\gamma$ kinematic map.
    This kinematic map suggests that, as already suspected, the low-velocity Br$\gamma$ (which is influenced by the jet) represent a distinct gas component.
    
    For the molecular phase (Fig.~\ref{fig:rot}, bottom panel), the line profiles are simpler, showing at most double peaked line profiles along the H$_{2}$ structure (blue contours). Each one of these peaks has distinct kinematics, one extended and the other reflecting the local molecular structure.
  
    Therefore, our kinematic maps represent the rotation of the molecular and ionized gas, with compatible PAs. But the kinematics related to the gas structures, seen by the emission line maps divided in velocity channels (Fig.~\ref{fig:h2}, right panel), are probably dynamically independent. In Appendix~\ref{sec:sk} we also explore the relation between the gas and stellar kinematics.
    
\begin{figure}
    \resizebox{\hsize}{!}{\includegraphics{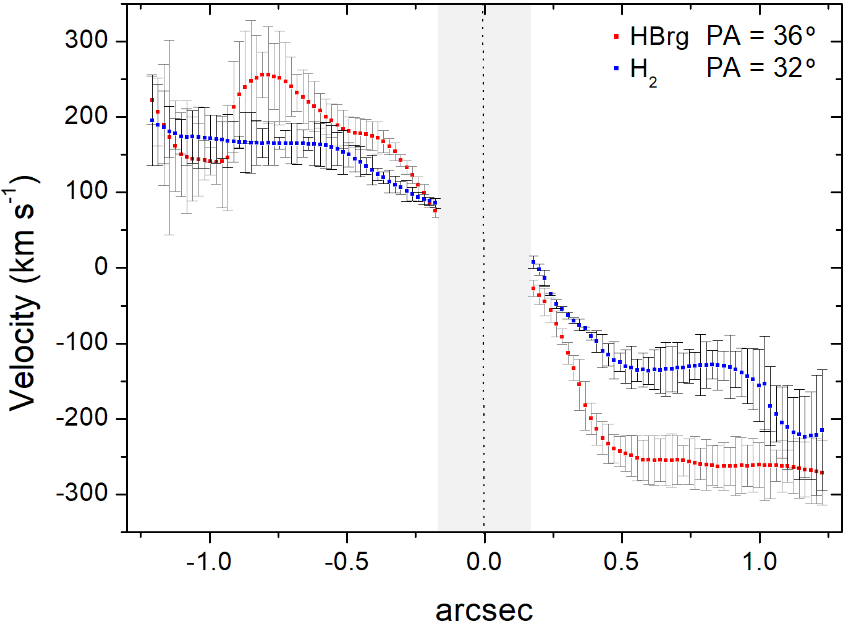}}
    \caption{Radial velocity curves for the H$_{2}$ $\lambda$21218 \AA~and the Br$\gamma$ emissions. The gray area corresponds to the masked nuclear region.}
     \label{fig:rotprofile}
    \end{figure}

    In Fig.~\ref{fig:rotprofile} the gas rotation curves are shown for the Br$\gamma$ and H$_{2}$ emission lines along their measured PA. Although the ionized gas presents velocities more than $\sim$100 km s$^{-1}$ larger than the molecular gas, one may see that the PA for the H$_{2}$ kinematic map does not cross its velocity peaks. 
 
    \subsection{The [Si\,{\sc vii}] coronal line emission}
    \label{sec:clb}
    
\begin{figure}
    \resizebox{\hsize}{!}{\includegraphics{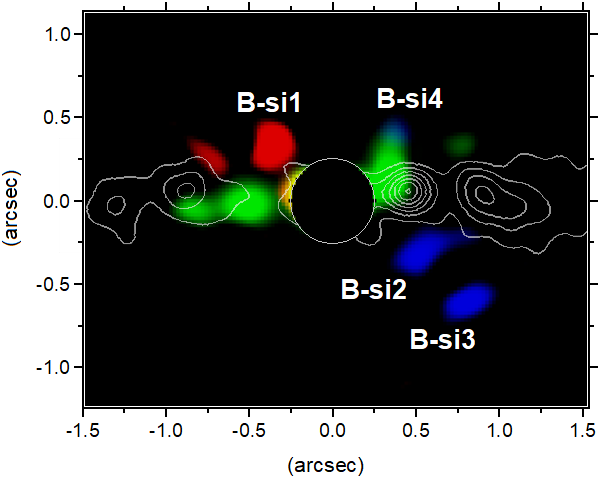}}
    \caption{RGB composition for the [Si\,{\sc vii}] $\lambda$24833 \AA~emission line overlaid with the radio emission (see velocity ranges in Table~\ref{table:vel}).}
     \label{fig:SiVII}
    \end{figure}

    The coronal line (CL) emission of [Si\,{\sc vii}] $\lambda$24833 \AA, with the highest ionization potential (IP) among the studied lines (205\,eV), was detected in the $K_{long}$-band and its spatial distribution is only shown through the velocity RGB map in Fig.~\ref{fig:SiVII} (see velocity ranges in Table~\ref{table:vel}), with the contours of the radio emission. A more detailed analysis will be presented by May D., et al. (in preparation). 
    The [Si\,{\sc vii}] distribution is mostly coincident with part of the Br10 structures, mainly for the low-velocity (see Fig.~\ref{fig:hbrk}). Such finding strengthen the hypothesis that this velocity regime is associated with the jet and the extended CL emission is excited by shocks along the jet orientation \citep{Ardila02,Ardila06,Ardila11,DMay18}.
    
    The most surprising feature, though, is that the high-velocity [Si\,{\sc vii}] does not seem to be related to the jet. However, it is worth noticing that here the measured velocities are much lower than the ones found for the galaxies NGC\,1068 and ESO\,428-G14 \citep{DMay18}. Such fact reduces the importance that is attributed to the jet in the gas cloud acceleration and excitation in NGC\,4151.
    
    We identified four coronal gas bullets (see Fig.~\ref{fig:SiVII}, with properties shown in Table.~\ref{table:bullets}). The bullet B-si4, however, calls attention because it represents the highest detected blueshift (of -664 \kms) among the CL bullets and is located outside the SW cone. In Sect.~\ref{sec:bullets} we will highlight their close relationship with the molecular structures.

    For the ESO\,428-G14 galaxy we found a highly ionized gas bubble in [Si\,{\sc vi}], with the brightest emitting region closely related to the jet.
    This is an evidence that, although the jet is not associated with the whole bubble structure, it still can be responsible for the bubble formation. Therefore, the detection of a cavity and a jet may motivate a closer look to their possible relation together with the CL emission. The hypothesis that NGC\,4151 could represent a similar case for the high-velocity CL emission - like the expanding cavities in ESO\,428-G14 - will be better explored in Sect.~\ref{sec:bubble}.
    
    \subsection{The X-ray emission}
    \label{sec:xray}
   
   \subsubsection{Correlation between the X-ray and the high-velocity [Fe\,{\sc ii}] emissions}
    \label{sec:xray}
  
     Observations from \textit{Chandra} of NGC\,4151 have been analyzed by \citet{Wang11a,Wang11b,Wang11c}, which discussed possible ionization mechanisms for the detected emission lines of [O\,{\sc vii}], [O\,{\sc viii}], [Ne\,{\sc ix}] and [Ne\,{\sc x}], in the soft X-rays. We will focus, however, on the [Ne\,{\sc ix}] emission, which shows in higher contrast the X-ray sub-structures and was already compared to NIR and radio data in \citet{Wang11b}, indicating good spatial correspondences between them. These authors argue that, for the most part of the NLR, the [O\,{\sc iii}]/X-ray line ratio points out to photoionization by the AGN, with an excess of X-ray emission in some locations where the gas seems to be shock-excited by the presence of the jet. Moreover, the PA for the soft X-ray emission (estimated from an elongated ``X''-shape structure in \citealt{Wang11a}) is 56\textdegree$\pm$6\textdegree, quite similar of what we found for the ionization cones in the NIR low-velocity [Fe\,{\sc ii}] emission (58\textdegree$\pm$3\textdegree), for a scale $\sim10$ times smaller.
     
\begin{figure}
  \resizebox{\hsize}{!}{\includegraphics{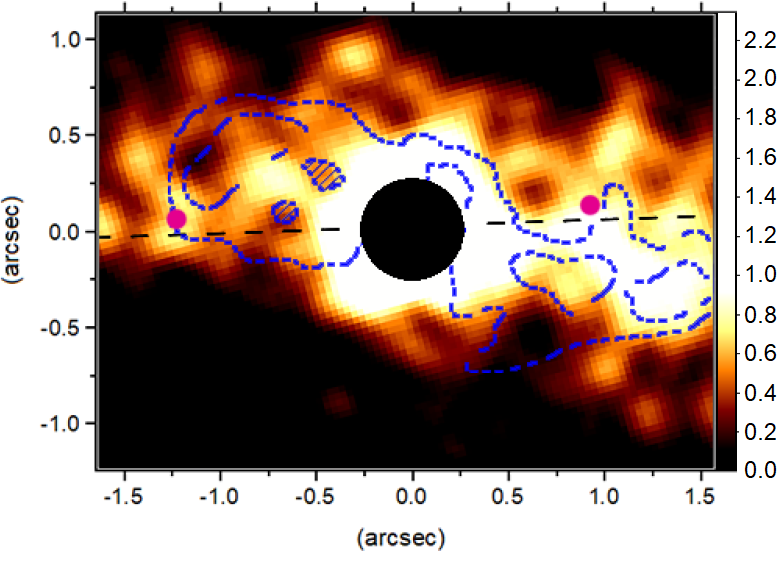}}
    \caption{Image of the soft X-ray emission line [Ne\,{\sc ix}] 0.24 keV overlaid with the high-velocity [Fe\,{\sc ii}] emission (blue contours). The jet PA is denoted by the black dashed line and the purple circles show where the [Ne\,{\sc ix}]/[O\,{\sc vii}] line ratios have the highest values. Flux units are in 10$^{7}$ photons s$^{-1}$ cm$^{-2}$.}
     \label{fig:xray}
    \end{figure}

     Since we reach a better spatial resolution in the NIR, more precise correlations may be found, and they are mainly expressed as two noticeable features: 1) there is no spatial correspondence between the [Ne\,{\sc ix}] and the high-velocity [Si\,{\sc vii}] emissions (namely, the [Si\,{\sc vii}] bullets), which is the studied NIR line with the highest IP; and 2) most of the [Ne\,{\sc ix}] distribution within the hourglass presents a good correspondence with the high-velocity [Fe\,{\sc ii}] emission (Fig.~\ref{fig:xray}).

    A correlation between the X-ray and the integral [Fe\,{\sc ii}] emission was already noted by \citet{Thaisa094151}, but here we found that the low-velocity part of the [Fe\,{\sc ii}] emission (Fig.~\ref{fig:Felh}, left panel), which is only bright at the borders of the hourglass, does not seem to have a spatial correspondence with the [Ne\,{\sc ix}] emission, bright along the cones' PA instead. The spatial coincidence between the high-velocity [Fe\,{\sc ii}] and the [Ne\,{\sc ix}] is unlikely to be a projection effect because the detected velocities for both emission lines are consistent with each other. \citet{Wang11b} estimated a velocity of 700 km s$^{-1}$ in the regions where the [Ne\,{\sc ix}] is probably excited by shocks, while we find de-projected velocities of $\sim$650 km s$^{-1}$ (considering $i=$45\textdegree~for the ionization cones - \citealt{Das05}) for the high-velocity [Fe\,{\sc ii}] emission (panels \textbf{d} to \textbf{f} in Fig.~\ref{fig:Febrvmaps}). 
    
    Looking at the [Ne\,{\sc ix}] and the high-velocity [Fe\,{\sc ii}] correlation, a natural question arises: how could a high-ionization specie, such as the [Ne\,{\sc ix}] (IP of 0.24 keV), be emitted from the same locations as those of the [Fe\,{\sc ii}] (IP of 7.9 eV)? The answer possibly resides in the fact that we are seeing an ionized shell structure unresolved by our data, caused either by photoionization or shocks \citep{Metzger14}. If it is due to shocks, some of the high [Ne\,{\sc ix}]/[O\,{\sc vii}] features \citep{Wang11b} could be explained by the continuous heating and destruction of dust grains right after the shock fronts, imprinting in the low-excited [Fe\,{\sc ii}] emission traces of these shocks, with correspondent velocities. Moreover, \citet{Colina15} also argue that the [Fe\,{\sc ii}] emission is indeed expected to be enhanced in presence of shocks, released by the destruction of dust grains.
    
    Therefore, we have found strong indications that the high-velocity [Fe\,{\sc ii}] and the [Ne\,{\sc ix}] emissions are physically related. Their relation with the jet, however, is more subtle. \citet{Wang11b} argue in favour of a connection between the jet and the X-ray emission at least in two regions of the NLR. We also confirm, with higher clarity, that the jet is likely responsible for the cones' asymmetry along the same direction. As shown in Sect.~\ref{sec:felh}, the high velocity [Fe\,{\sc ii}] bullets are on the borders of the radio knots \textbf{C2} ad \textbf{C5}. In this sense, a fraction of the high-velocity gas still may be under the influence of the jet at its borders as much as the enhanced [Ne\,{\sc ix}] emission. The low-velocity [Fe\,{\sc ii}], in turn, is clearly enhanced in the walls of the hourglass closer to the jet, as well as all the low-velocity ionized gas emission.   
    
   \subsubsection{Correlation between the X-ray and molecular emissions}
    \label{sec:xrayh2}
    
    \citet{Wang11a} found an excess of hard X-rays compared to the soft X-ray emission (hardness ratio) in the nuclear region along an orientation compatible with the H$_{2}$ distribution outside the cones, suggesting a higher obscuration in the AGN equatorial region, as expected. 
    This result also shows that the central radiation is not totally blocked by a supposed clumpy torus in NGC\,4151, which rises a curious question: the morphology of the molecular gas (resembling a double shell outside the cones) could be related to the passage of some variable fronts of X-ray emission centred in the nucleus? Such morphology would be the response of the nuclear molecular gas heating due to variations events in the X-ray emission, imprinting a bright shell for each X-ray ``pulse''. The distance of the H$_{2}$ NW arcs to the nucleus (segments \textbf{1A} and \textbf{2B} in Fig.~\ref{h2}; right panel) are $\sim$100 and 180 ly, respectively. The shortest interval is, at least, comparable to the beginning of the historical records of optical variability in NGC\,4151 \citep{Ok16}. If we can associate the optical to X-ray variability in NGC\,4151 \citep{Ok94}, no secular variation has ever been found and, therefore, there is no possibility to associate the H$_{2}$ morphology with an enhancement due a long-term X-ray variability.
  
    On the other hand, the highest values of the [Ne\,{\sc ix}]/[O\,{\sc vii}] ratio seem to be spatially related to the external edges of the molecular segments that are in the process of fragmentation (purple circles in Fig.~\ref{fig:h2}). Those peaks are located on the radio knots \textbf{C2} and \textbf{C6} and on the [Fe\,{\sc ii}] bullets (Fig.~\ref{fig:Febrvmaps}, panels \textbf{h} and \textbf{i}).
    We have found for the galaxies NGC\,1068 (M\&S17) and ESO\,428-G14 \citep{DMay18} that the region where the bullets arise are associated with locations of strong H$_{2}$-jet interaction. In the case of the galaxy ESO\,428-G14, there are also two extra-nuclear X-ray peaks related to a pair of [Si\,{\sc vi}] bullets. Such signs of highly energetic processes suggest that the interaction between the central source and the molecular gas may have a central role in the NLR dynamics.

    \section{Discussion}
    \label{sec:discussion}
    
     \subsection{The misalignment between the torus and the accretion disc}
    \label{sec:desalin}
    
    The NLR shape in NGC\,4151 differs from the classical ionization cones, without a well defined vertex and having a more linear structure, with a narrow opening angle for larger distances \citep{Perez89,Pedlar93}. To reconcile these features with the Unified Model, \citet{Robinson94} proposed a geometry where a cone with a wide opening angle intercepts the galactic disc (their Fig.~20). In this scenario the jet is aligned with the cones axis and its apparent misalignment with the NLR would be a projection effect.
    Since this model was proposed, for a NLR extension up to 40 arcsec (more than 2.5\,kpc), it continues to be adopted until now (e.g. \citealt{Das05,Thaisa104151}), which may not be the case for scales down to few arcsec .
    One point that seems to have gone unnoticed is that it is hard to explain the highest detected outflow velocities (up to -1700 \kms; \citealt{Winge97,Kaiser00,Das05}, at distances $\gtrsim$50 pc from the nucleus) if the NLR is a consequence of the disc intercepted by the ionization cones. Such values would represent much higher de-projected velocities than the expected for the outflowing gas in the intercepted disc (v$\sim$3000 \kms for an inclination of $\sim$20\textdegree), which are similar to the ones found for the BLR of this galaxy \citep{Arav98}.
  
    According to the results presented in Sect.~\ref{sec:felh}, the low-velocity [Fe\,{\sc ii}] depicts the glowing walls of an hourglass structure, which is equivalent to the walls of the ionization cones. Geometrically speaking, the emitting walls of the ionization cones, projected in the plane of the sky, are detected where there is more emission along the line of sight (LoS). However, in the current scenario, there are no ``cones walls'' accumulating more emission along the LoS, because these walls would be outside the intercepted disc. In this way, there would be no preference for a low-velocity emission of [Fe\,{\sc ii}] along any edge, but only a trend to see a uniform emission coming from the cones interior, as modelled by \citet{Stalevski17}. Furthermore, since we are seeing the cones walls for a scale less than 3 arcsec ($\sim$150\,pc), we have another argument in favour of our interpretation: here we are necessarily seeing all the emission embedded within the galactic disc, so there is no partial intersection; therefore, the detected hourglass walls certainly represent the partially ionized gas shining in the real edges of the cones. 
    Based on these arguments we change the current interpretation and favour the scenario where the misalignment between the jet and the ionization cones is real, instead of a projection effect. 
    
    The narrow appearance of the ionization cones seen for the kiloparsec scale in \citet{Robinson94} seem to be a natural extension of the hourglass narrow opening angle seen here. In short, we are saying that the narrow appearance of the ionization cones in \citet{Robinson94} is not justified by a small fraction of galactic disc intercepted by a larger cone. The narrow opening angle is, in fact, the real shape of the cones and they are misaligned with respect to the jet.
    In fact, we may speculate that the ionization cones could be narrower than expected due to the abundant presence of extended H$_{2}$ emission right after the the cones walls, what may play a role in collimating the central radiation in this case.
    Since the radial acceleration of the outflow modelled by \citet{Das05} has no influence from the jet and we have indications that the jet only disturbs the gas in the systemic-velocity, in principle there is no inconsistency between their model and our new claim of the real misalignment between the jet and the cones.
   
    In addition, there is no strong reason or statistical basis to advocate that the axis of the jet and the ionization cones should be the same (in fact, \citealt{Pedlar93} suggested that the jet has a PA=40\textdegree~with respect to the LoS). In NGC\,1068, for instance, the jet (before bending) is misaligned by 23\textdegree~with respect to the ionization cones (also depicted by an hourglass structure seen in the low-velocity [Fe\,{\sc ii}] emission).
    In NGC\,4151, this misalignment is similar, of $\sim$20\textdegree, but here it is more evident because the jet maintains its orientation along the NLR (in NGC\,1068 the jet bends to the same orientation of the cones). Such a fact implies that the torus and the accretion disc have their axis misaligned by, at least, the same value.
    Therefore, since the extended NLR \citep{Robinson94} present the same PA of that of the ionization cones showed by our FoV, the outflow showed here is probably expelled out of the galaxy disc on larger scales.
    
    \subsection{The relation between the H$_{2}$ and [Fe\,{\sc ii}] emissions}
    \label{sec:arc}
    
\begin{figure*}
    \resizebox{\hsize}{!}{\includegraphics{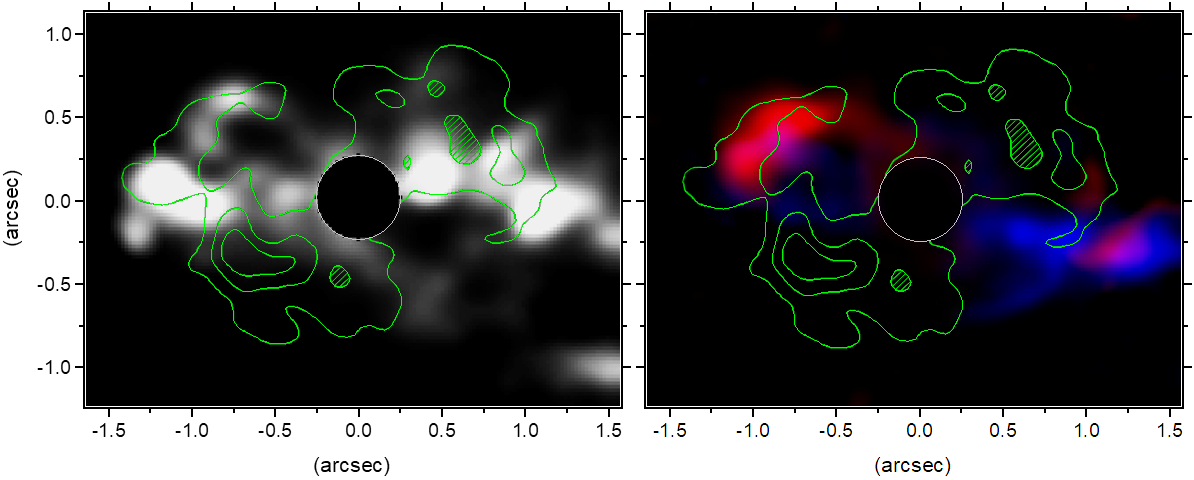}}
    \caption{The low- and high-velocity [Fe\,{\sc ii}] emissions from Fig.~\ref{sec:felh}. The velocity range is shown in Table~\ref{table:vel} and the contours denote the integrated H$_{2}$ emission from Fig.~\ref{fig:h2} (left panel).}
     \label{fig:3f}
    \end{figure*}

    In Fig.~\ref{fig:3f} we show how the low- and high-velocity regimes for the [Fe\,{\sc ii}] emission are related to the molecular gas (in contours). The first thing to notice is the H$_{2}$ emission in the extremities of the NE and SW cones (with velocity ranges between 100 $\lesssim |v|\lesssim$300~km s$^{-1}$), which seem both spatially and kinematically consistent with some [Fe\,{\sc ii}] structures (mainly for $|v|\lesssim$400 km s$^{-1}$). This relation becomes clearer in the BRV maps of Fig.~\ref{fig:Febrvmaps} (panel \textbf{c}), that show a series of [Fe\,{\sc ii}] arcs possibly connected to some identified segments of H$_{2}$ in Fig.~\ref{fig:h2} (right panel) (namely, the association [arc1 - 1C], [arc2 - 2B] and [arc3 - 2C]).
    
    Our hypothesis for segment \textbf{2C} (Fig.~\ref{fig:h2}, right panel) is that it would be part of an already dissociated dusty and molecular structure formerly contained between segments \textbf{2A-2B}. Furthermore, the molecular segment \textbf{1C} (which is still emitting) would be once part of segment \textbf{1A-1B} in the NE cone.
    In short, according to this interpretation, segments \textbf{2C} and \textbf{1C} were before filling the holes between \textbf{2A} and \textbf{2B} and between \textbf{1A} and \textbf{1B}, respectively, in the form of molecular gas. Now they are being dissociated and pushed outwards by the outflow and/or AGN radiation.
    In addition to the relationship between the [Fe\,{\sc ii}] arcs and the H$_{2}$ segments, there is a good correspondence between two low-velocity lobes detected for the same lines, which can be seen in panels \textbf{a} of Figs.~\ref{fig:H2brvmaps} and \ref{fig:Febrvmaps}.
    Therefore, the molecular and [Fe\,{\sc ii}] emissions are highly suggestive of a prior H$_{2}$ structure around the nucleus that still may have its symmetry recovered from these two distinct gas phases, complementing each other. The identification of the molecular segments presented in Fig.~\ref{fig:h2} (right panel) is probably the most simple arrangement for the observed H$_{2}$ emission in form of an ``unified'' molecular structure that is partially destroyed at the present stage.
    
    Since the cones are not a projection effect of a larger cone intercepting the galactic disc (as discussed in the previous section) and the [Fe\,{\sc ii}] is dynamically connected to the molecular gas, both should be part of the same outflow event in a direction outward the galactic plane (despite their emission are embedded into the galactic disc in the present FoV). In this context, there are still two scenarios for the H$_{2}$ emission at the shadow of the cones, with (1) - the most natural one and more favoured by the literature (as well as for others AGN - \citealt{Muller09,Thaisa09,Davies09,Davies14}) - saying that the H$_{2}$ represents the gas in the disc feeding the AGN or, (2) that its detected structure may be entirely part of an outflow.
    
    Some observational conditions to assume inflow of gas are well described in Sect.~3.2.4 of \citet{Shimizu19}, which detect inflow in NGC\,5728 (see also the review of \citealt{Thaisa19}). In short, one should detect non-rotating molecular gas - in redshift - in the near side of the galaxy disc (toward the nucleus) and - in blueshift - in the far side (also flowing to the nucleus). Even with just this condition, this case does not apply to the H$_{2}$ emission in NGC\,4151, as already noted by \citet{Thaisa104151}. In fact, according to this condition, the H$_{2}$ kinematics next to the nucleus points out to molecular outflow also outside the cones (Fig.~\ref{fig:h2}, right panel).
    
    Also in favour of the ``feeding scenario'', there is the possibility that the H$_{2}$ emission close to the systemic velocity could be slowly feeding the AGN at velocities lower than our detection limit. 
    But, in a rotating disc intercepted by the outflow, from the moment when the molecular material is exposed to the ionization cones it would begin to be pushed away by the central source. As a consequence, we may expect to have a discontinuity along the disc because there will be no material going out at the other side of the cone (and at the same distance from the AGN).
    In other words, the molecular gas entering one cone would be pushed away by the outflow and would not ``leave'' at the other edge, interrupting the observed pattern.
    
    To check if this prediction leads to a detectable ``gap'' we need to know the H$_{2}$ rotation velocity and the fraction of the outflow time scale after interacting with the molecular structure. For the rotation velocity we adopted a conservative value of 200 \kms for the H$_{2}$ (without correcting for the disc inclination - Fig.~\ref{fig:rot}). For the distance, we find outflowing gas at least 0.75 arcsec (50\,pc) from the SW H$_{2}$ segments. According to panel \textbf{b} of Fig.~\ref{fig:Febrvmaps} there is [Fe\,{\sc ii}] in outflow at 285 \kms for the same distance (corrected for the cone inclination of 45\textdegree). These values represent $1.7\times10^{5}$\,yr of disc rotation since the first contact between the outflow and the SW molecular segment, resulting in a displacement (or a ``gap'') of 40\,pc between the segments and the outer cones' walls.  
    
    Taking into account the rotating direction and disc inclination, this means that segments \textbf{1A} and \textbf{2A} should be more than 40\,pc distant from the cones edges (what would be detectable at our resolution), but no gap is seen. 
    In other words, with the presence of the ionized outflow in the cones’ direction, a given rotating H$_{2}$ segment (e.g. \textbf{1A} to \textbf{1B}) should follow rotation after the outflow destroyed its \textbf{1A}-to-\textbf{1B} bridge and should thus show a spatial gap from the cones’ wall to where the segment \textbf{1A} restarts, what is not seen.
    One may think that the segment could leave the cone's interior more or less intact (leading to no gap at all), but we already do not see a continuous structure between segments \textbf{2A} and \textbf{2B}.
    There is another indication that the H$_{2}$ structure is not part of a rotating disc intercepted by the outflow: in a rotating disc the segment \textbf{1B} would be entering at the NE cone and, as trend while it rotates, it would be continuously farther from the AGN as the outflow pushes it outwards, but we see exactly the opposite.
    
    Therefore, considering our lower limit for the relevant outflow time scale (what would possibly increase for a lager FoV), the scenario of a rotating H$_{2}$ structure is not favoured by our data.
    On the other hand, if the entire H$_{2}$ emission is expanding, we expect to see exactly what is observed, with the structure outside the cones presenting nearly systemic velocities (as seen in projection in the plane of the sky). NGC\,1068 is a clearer example of an expanding cavity, with exactly the same behaviour (M\&S17).
    
    We agree that the intuitive interpretation for the low-velocity H$_{2}$ structure should be the inflow of gas, since the AGN feedback occurs in the cones and the gas feeding in the cones' shadow. However, there is no evidence for this interpretation in NGC\,4151 and one expects a much lower gas fraction (of the calculated 429\,\Msun~- Sect.~\ref{sec:h2}) to feed the AGN, as suggested by the low Eddington ratio of $\sim$1 per cent.
    In this work, we provide a possible and alternative interpretation of an outflowing H$_{2}$ structure, setting a ``new problem'' for the feeding-feedback dynamics in this galaxy, which requires further investigation. Although more evidence has to be found, in the following section we present a stronger argument in favour of this scenario with the analysis of a set of compact clouds identified along the H$_{2}$ segments.

     \subsection{Molecular gas cloud fragmentation: on the origin of the bullets}
    \label{sec:bullets}
   
    This section relates the molecular and ionized gas emissions for the H, [Fe\,{\sc ii}] and [Si\,{\sc vii}] lines.
    As shown in the previous section, the H$_{2}$ structures inside the ionization cones (both for the low- and high-velocity regimes) seem to be dynamically related to the low-ionization [Fe\,{\sc ii}] emission line. The low-velocity molecular gas outside the cones also seems to be complementary with respect to the low- and high-velocity [Fe\,{\sc ii}] (Fig.~\ref{fig:3f}), which is clearly seen in the case of NGC\,1068 (M\&S17, their Fig.22).
    Moreover, the general picture displayed in Fig.~\ref{fig:hbr_dual} (left panel) and Fig.\ref{fig:SiVII} shows that the low-velocity ionized gas (which is part of the hourglass walls) seems to be related to the jet (mainly for the Br$\gamma$ and [Si\,{\sc vii}] emission lines). The high-velocity gas, in turn, shows more fragmented structures (inside the hourglass) for all the cases. We were able to identify, for instance, candidates of symmetrical gas bullets (with respect to the nucleus), from where a closer relationship between the emission lines will be explored.

\begin{figure*}
    \resizebox{0.85\hsize}{!}{\includegraphics{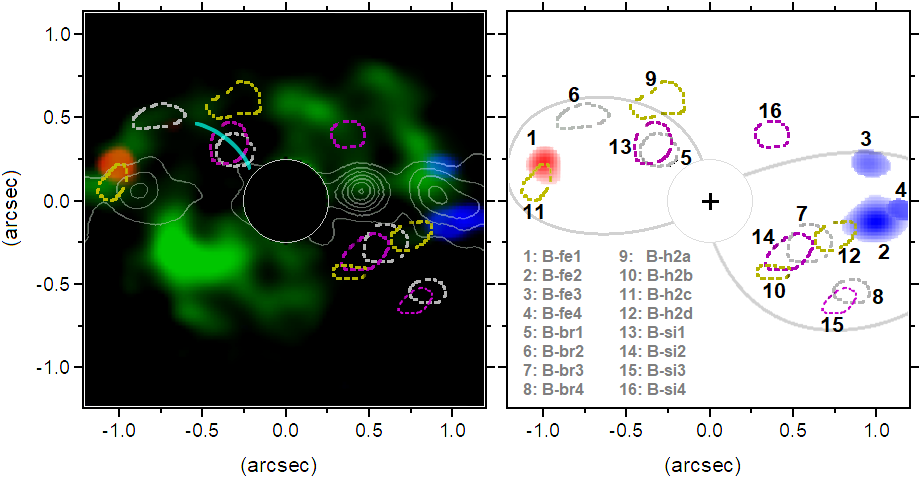}}
    \caption{Left panel: image of the H$_{2}$ $\lambda$17480 \AA~line (green) with the gas bullets of [Fe\,{\sc ii}] (blueshifted/redshifted blobs, from Fig.~\ref{fig:Febrvmaps}), [Si\,{\sc vii}] (purple contours, from Fig.~\ref{fig:SiVII}), Br10 (gray dashed contours, from Fig.~\ref{fig:Hbrbrvmaps}) and H$_{2}$ (yellow contours, from Fig.~\ref{fig:H2brvmaps}). All the bullets follow the same redshift and blueshift trend as those of [Fe\,{\sc ii}], at their cone's sides. The radio emission is shown in white contours and part of the H$_{2}$-lobe (seen in Fig.~\ref{fig:H2brvmaps}, panel \textbf{a}) is represented by the pale blue line. Right panel: bullets identification and the hourglass shape for a more straightforward visualization.}
     \label{fig:bullets}
    \end{figure*}

    In Fig.~\ref{fig:bullets} (left panel) we show the integral H$_{2}$ distribution (in green) with the radio emission contours (white). The gas bullets, selected from the BRV maps, are displayed in red and blue colours for the [Fe\,{\sc ii}] emission and in contours for the other lines (see figure caption), with their identifications shown in Fig.~\ref{fig:bullets} (right panel) and properties listed in Table~\ref{table:bullets}.
    The bullets are quite compact, with typical sizes between 10 to 20\,pc, which is slightly above the FWHM of the PSF.
    
    One may clearly see that the bullets are not randomly distributed but, instead, they seem to complement the molecular emission, filling the gaps between segments \textbf{1A-1C} and \textbf{2A-2B} shown in Fig.~\ref{fig:h2} (right panel). The close pair of Br10 and [Si\,{\sc vii}] SW bullets would be related, in turn, to the supposedly extinct molecular segment \textbf{2C}, which is the symmetrical counterpart of segment \textbf{1C} (in the NE cone) still in process of fragmentation. 
    
    On the other hand, the bullets B-br1 and B-si1 seem to present a distinct behaviour, with the lowest detected velocities and not related to any molecular segment. However, they are located exactly over the lobes' walls seen both for the H$_{2}$ and [Fe\,{\sc ii}] emissions (panels \textbf{a} of Figs.~\ref{fig:H2brvmaps} and \ref{fig:Febrvmaps}), a structure that has no counterpart in the SW cone. Therefore, all the bullets (even the outlier bullets B-br1 and B-si1) would be associated with the fragmentation of a molecular structure, at different stages. 
    
    It is interesting to note the bullet B-si4, a high-velocity cloud outside the hourglass structure and, apparently, not directly exposed to the central source. This bullet seems to be spatially associated with the molecular segment \textbf{1A} in a region close to the systemic velocity. Although the nature of this bullet is uncertain, it is located above the radio knot \textbf{C3}, where the [Fe\,{\sc ii}] emission shows a curious structure at the shadow of the SW cone (Figs.~\ref{fig:FeII}). The presence of structured features outside the ionization cones are likely consequences of localized process in the NLR, such as jet-cloud interactions (see more details in Sect.~\ref{sec:comp}).

    With the exception of the outlier bullets, their projected velocities follow a similar trend in both cones: for the H$_{2}$ bullets we found an average value of $\sim$200 km s$^{-1}$, for the Br10 and [Si\,{\sc vii}] (inside the cones) $\sim$360 km s$^{-1}$ and for the [Fe\,{\sc ii}] we have the highest velocities, up to $\sim$2000 km s$^{-1}$.
    
    The [Fe\,{\sc ii}] bullets, as the farthest ones, also call attention because of their high degree of symmetry with respect to the nucleus (like B-fe1 and B-fe2 bullets, with virtually the same distance and orientation). All of them are also close to the radio knots \textbf{C5} and \textbf{C2}. Since the bullet B-fe4 (right next to B-fe2) has a difference in absolute velocity of $\sim$900 km s$^{-1}$ to the bullet B-fe1, it would be highly unlikely that we were witnessing the exact moment in which both bullets overlap two opposite molecular segments. However, if the bullets are associated with their respective molecular segments, which have velocities bellow $\sim$200 km s$^{-1}$, we need to explain how the bullets and the molecular segments still maintain nearly the same distance from the nucleus. The most reasonable explanation is that a more dramatic event is taking place where the jet interacts with the molecular gas (close to the [Fe\,{\sc ii}] bullets), accelerating the bullets to larger projected velocities. In NGC\,1068, where the fragmentation of molecular walls is more evident, one may easily see that the [Fe\,{\sc ii}] and [Si\,{\sc vi}] compact clouds reach velocities up to 4 times larger than the H$_{2}$ structures from where they are located.
    
    On the other hand, the low-velocity emission along the jet (as seen for the Br10 emission in Fig.~\ref{fig:hbr_dual}, left panel) would be naturally explained by a less dramatic lateral expansion of the gas, without any strong interaction with the ISM. This fact leads us to the conclusion that the low-velocity gas and the bullets (high-velocity) are influenced by the jet and follow the same orientation within the ionization cones.
   
    Apart from the [Fe\,{\sc ii}] bullets, the velocity of the other bullets are just a little larger than that measured for the fragmented H$_{2}$ segments. This evidence suggest that we are mostly seeing a recent process of fragmentation, with the highest detected velocities where the jet interacts with the molecular segments.
    
    The velocity dispersion has an average value of 185 km s$^{-1}$ for the [Fe\,{\sc ii}] bullets, 59 km s$^{-1}$ for the [Si\,{\sc vii}] and 85 km s$^{-1}$ for the H$_{2}$. For NGC\,1068 we have found 152, 112 and 95 km s$^{-1}$, respectively. Despite the small number of clouds, a low velocity dispersion for the coronal blobs in NGC\,1068 is also seen for those that are close to the molecular arm under fragmentation (68 km s$^{-1}$). Therefore, the same trend for the velocity dispersion seen in the compact clouds of NGC\,1068 is found for the gas bullets in NGC\,4151.
    
    In essence, we are offering a possible explanation for the outflow in NGC\,4151, connecting the molecular and ionized gas phases in light of the results found for NGC\,1068. 
    In this scenario we argue that the outflowing gas bullets are originated from the fragmentation of some molecular segments exposed to the AGN.
    The origin of the most dramatic ones - the high-velocity [Fe\,{\sc ii}] bullets - seems to be directly affected by the jet. These regions are highly suggestive of the presence of shocks, expanding even further the molecular segments associated with the [Fe\,{\sc ii}] arc-shaped structures (Fig.~\ref{fig:Febrvmaps}, panel \textbf{c} and the main structures seen in Fig.~\ref{fig:FeII}).
    
    The detection of all bullets and their relation to the molecular segments, both within and outside the cones, bring the strongest indication that the H$_{2}$ emitting structure is part of the outflowing gas. First, we should remember from Sect.~\ref{sec:desalin} that the low-velocity [Fe\,{\sc ii}] emission depicts the real cones' shape, changing the current paradigm of a disc intercepted by a cone with large opening angle. Even if one may think that within 50\,pc the cones could be interacting with the disc, projected velocities above 1500 \kms (as the bullets B-fe3 and B-fe4) are too high for a disc inclined by just 21\textdegree~\citep{Pedlar92} (what would result in outflow velocities of $>$4000 \kms for the same distances).
    
    Thus, the outflowing bullets are moving in a direction out of the galactic plane (like those of [Fe\,{\sc ii}] in segments \textbf{1C} and \textbf{2B}). Since the bullets are clearly associated with the molecular gas fragmentation, inside and outside the cones (and this is observed), it is hard to sustain the hypothesis that the H$_{2}$ emitting structure is not part of the outflow as well.
    
    The low velocities for the H$_{2}$ structures outside the cones are expected if we are seeing molecular walls of an expanding cavity. This is exactly the same effect explained for the low-velocity [Fe\,{\sc ii}] emission in the hourglass walls (Sect.~\ref{sec:desalin}), where - in a 3D hollow structure - we naturally see its projection in the plane of the sky at low velocities. Therefore, we favour the hypothesis where both gas phases (molecular and ionized) take part of the same outflow event.
    
    \subsubsection{Outflow energetics}
    \label{sec:kl}
    
    We may starting to estimate the kinetic luminosity of the Br10 bullets measuring the flux, mass and dynamical time scale to reach the current position, for each bullet, from the equations 
    
    \begin{equation}
    \begin{aligned}
    M_{Br10}&=0.11~\textit{f}~\left(\frac{L_{Br10}}{erg s^{-1}}\right)\left(\frac{n_{e}}{10^{4}~ cm^{-3}}\right)^{-1}~\textrm{g}
    \label{h2mass}
    \end{aligned}
    \end{equation}
    
     \begin{equation}
    \begin{aligned}
    \dot{E}\approx\frac{\dot{M}}{2}(v^{2}+\sigma^{2})
    \label{h2}
    \end{aligned}
    \end{equation}

    \noindent where \textit{f} is the filling factor and $n_{e}$ the electronic density. The emission coefficient of the Br10 recombination line was extracted from \citet{Osterbrock06}. Assuming \textit{f}=0.1 and $n_{e}=10^{4}$ cm$^{-3}$ for the bullets, the resulting masses are 37$\Msun$, 14$\Msun$, 49$\Msun$ and 29$\Msun$ for the B-br1, B-br2, B-br3 and B-br4 bullets, respectively. Using the projected velocities shown in Table~\ref{table:bullets}, the dynamical time scale for each one is $1.9\times10^{5}$, $1.4\times10^{5}$, $1.2\times10^{5}$, and $2.0\times10^{5}$ years, respectively. From these values and the velocity dispersion $\sigma$, the kinetic luminosity $\dot{E}$ of each bullet is estimated to be $2.0\times10^{36}$, $5,9\times10^{36}$, $1.7\times10^{37}$ and $5,0\times10^{36}$ erg s$^{-1}$. Since the bolometric luminosity of NGC\,4151 is $7.4\times10^{43}$ erg s$^{-1}$ \citep{Crenshaw15}, the four bullets represent only 4$\times10^{-5}$~per cent of the total radiated energy for this galaxy.
    
    If we take into account all the high-velocity Br10 emission, with a calculated mass of $10^{4}\Msun$, we increase by a factor of $\sim$60 this contribution to the kinetic luminosity.
    The remaining 8 ionized bullets of [Fe\,{\sc ii}] and [Si\,{\sc vii}] may be thought as different ionization phases of the bullets during the outflow, with a similar kinetic luminosity to that of the Br10. When we add to this equation the 4 molecular bullets (which sum $1.7\times10^{36}$ erg s$^{-1}$) and all the high-velocity Br10 emission, the kinetic luminosity reaches $\lesssim$0.01 per cent of the total AGN luminosity. Even assuming that all the H$_{2}$ emission is in outflow (with v$\sim$100 \kms), the kinetic luminosity is only $\sim2\times10^{37}$ erg s$^{-1}$, comparable to the 4 Br10 bullets.
    
    Our estimate of the kinetic luminosity (limited to the ionized and warm gas phases) is at least two orders of magnitude lower than the low-end of the range found for a few nearby AGN \citep{Crenshaw12} and bellow the threshold for an efficient AGN feedback (according to predictions of \citealt{Kurosawa09,Hopkins10}).
    However, since it is unlikely that 100 per cent of the energy injected by the AGN into the ISM will become kinetic power of the outflow, the threshold for an efficient feedback mechanism is probably lower than that given by most theoretical models.
    Moreover, comparing the high values of kinetic luminosity given by theory with the observed ones may be tricky also because we are not seeing all the outflow phases and it takes time to the outflow to have a larger impact through the galaxy \citep{Harrison18,Cicone18}.
    
    \subsection{Molecular arms or cavity walls?}
    \label{sec:bubble}
    
    The molecular segments named in Fig.~\ref{fig:h2} (right panel) could be naturally interpreted as molecular arms in the galactic plane, but the sustained scenario, where the molecular structures are related to the origin of the ionized bullets (which traces the outflow), deviates from this interpretation. At the end of Sects.~\ref{sec:arc} and ~\ref{sec:bullets} we discussed that, even close to the systemic velocity, the H$_{2}$ at the shadow of the cones cannot be seen as a dynamically distinct structure from those exposed to the AGN. If settled in the disc before the outflow event, the H$_{2}$ would be rotating (Sect.~\ref{sec:kin}) and, as a consequence, the molecular segments outside the cones would not be close to the cones' walls at the present stage of the outflow. 
    The only difference between the cones is that, while the NE molecular segment is under fragmentation, the SW one was already destroyed.
    
    As discussed in Sect.~\ref{sec:arc}, the most evident relation between the [Fe\,{\sc ii}] and molecular emissions is shown for lower [Fe\,{\sc ii}] velocities, as the Fe-lobe and the [Fe\,{\sc ii}] structure arc1 (Fig.~\ref{fig:Febrvmaps}, panels \textbf{a} and \textbf{c}, respectively), which present molecular counterparts both spatially and in velocity. In this way, the molecular segment \textbf{1C} (Fig.~\ref{fig:h2}, right panel) is most likely part of the outflowing gas.
    
    Furthermore, there are the molecular and ionized bullets B-si4 and B-h2a right after the cones' edges, with the former located in segment \textbf{1A} and the latter between segments \textbf{1A} and \textbf{1C}.
    In NGC\,1068 there is also a compact H$_{2}$ cloud clearly outside the cones (panel \textbf{a} of Fig.19 and panel \textbf{b} of Fig.22 in M\&S17) and, there, we argued that the whole H$_{2}$ structure was more compact when the fragmentation process began. In this way, some regions now outside the cones were before exposed to the AGN.
    Although an exception, these outer bullets should have a similar origin from those within the cones. Therefore, we have as observational evidence that the outflowing bullets are always associated with H$_{2}$ structures which, in turn, are complementary with each other around the AGN and resembles a cavity (Fig.~\ref{fig:H2brvmaps}, panels \textbf{b} and \textbf{c}). 
    
    In other words, all the 16 bullets seem to be associated with the molecular segments; so, if this association is real, it would not be possible that the segments are also accelerated outwards? In this case we offer an alternative explanation rather than assuming the H$_{2}$ segments (or ``arms'') as part of the AGN feeding and the bullets as part of the feedback.
    It is worth mentioning that this interpretation is not incompatible with the AGN feeding, which can occur on much smaller scales.
    
    The apparent discrepancy between the velocities of the molecular segments and the bullets is expected in case of an expanding cavity because we see only the cavity walls with projected velocities close to zero (it could be and probably is expanding faster in the plane of the sky). The highest [Fe\,{\sc ii}] velocities, for instance, could be also due to the lateral expansion of the gas close to the jet, as indicated by the presence of shocks in Sect.~\ref{sec:xrayh2}.

    In the case of NGC\,1068, the connection between the H$_{2}$ emission inside and outside the cones is much more evident, with the wall exposed to the AGN (in the NE cone) also being fragmented into compact blobs of ionized gas. This wall is, in fact, part of a molecular cavity surrounding the AGN, with a complementary fraction of H$_{2}$ outside the cones following the same round structure.
    There, we have shown that the H$_{2}$ surrounding a cavity is an expanding molecular bubble being disrupted by the jet in the NE cone and it is still preserved in the SW cone.
    Therefore, giving the similarity between the scenario of molecular gas fragmentation for the galaxies NGC\,4151 and NGC\,1068, it is natural to consider the possibility that the H$_{2}$ segments in NGC\,4151 would represent the molecular walls of a more complex cavity structure.
    
    In NGC\,1068 we are seeing the NLR almost at the plane of the sky (i=5\textdegree~- \citealt{Crenshaw00,Fischer13}) and here, at 45\textdegree, an asymmetric cavity would not be projected in the same obvious way. Moreover, the misalignment between the cones and the jet may also contribute to the complexity of the cavity structures. Our attempt to unify the H$_{2}$ segments into a symmetrical structure (Fig.~\ref{fig:h2}, right panel) may be close to what this cavity system could look like, in projection. The main difference between the dynamics presented by the cavities in the two galaxies is that in NGC\,4151 both sides of the cavity are in process of fragmentation, with one in a more advanced stage (SW side) than the other (NE side). The segments that represent this process (\textbf{1C} and \textbf{2C}) are the most distant from the AGN, possibly pushed away by radiation pressure from the AGN. Moreover, the tips of these segments that coincide with the jet orientation are even more distant, suggesting that the jet may have a role in shaping the cavity.
    In NGC\,1068, in turn, the SW side is inflating without signs of fragmentation.
   
\begin{figure}
  \resizebox{\hsize}{!}{\includegraphics{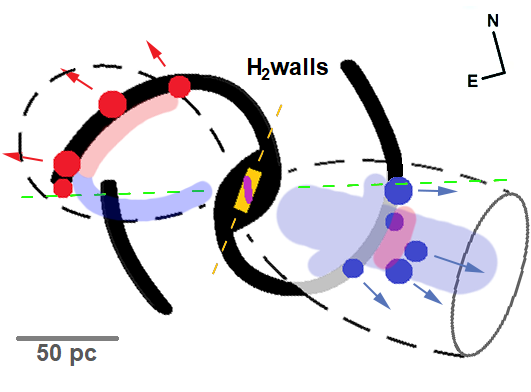}}
    \caption{Sketch of the scenario proposed for the NLR of NGC\,4151. The discontinuous H$_{2}$ walls are shown in black together with the dashed contours of the hourglass structure. The filled circles represent the bullets of ionized and molecular gas in redshift and blueshift, with the arrows denoting their radial motion, while the pale-blue and red colours illustrate the main regions of the high-velocity [Fe\,{\sc ii}] emission. The straight dashed lines denote the PA of the jet (green) and the torus (orange), while the accretion disc is shown in purple (not to scale).}
     \label{fig:model}
    \end{figure}

    In Fig.~\ref{fig:model} we show a scheme that illustrates the scenario proposed for NGC\,4151. There, the molecular gas is shown in black, with the NE wall in process of fragmentation and the SW one already destroyed, with both walls resulting in bullets of molecular and ionized gas. These walls are not in the galactic plane. The bicone is with an open side, but our FoV does not allow to confirm whether this is the case.
    The cavity scenario for NGC\,4151 is consistent with the detected correlations between the molecular and ionized gas phases, and is also inspired by the model presented for NGC\,1068 (upper panel of Fig.30 in M\&S17). Here, we are presenting possibly the simplest molecular configuration to explain such correlations. The scheme simplifies the [Fe\,{\sc ii}] high-velocity emission from Fig.~\ref{fig:Felh} (right panel) through the light red and light blue colours, mostly covering the inner molecular walls. Finally, the misalignment between the torus and the accretion disc (orthogonal to the jet) is highlighted in the centre, out of scale.    
    
    Since the molecular gas, as part of the outflow, is in expansion (with the projected walls naturally seen in low-velocity), the molecular reservoir was once more concentrated in the nucleus. We have suggested that the misalignment between the axis of the torus and the accretion disc may continuously eject a huge amount of gas due the interaction between the jet/radiative wind with a torus/molecular disc (\citealt{Marscher06}, \citealt{Kawakatu09} and individual cases of 3C\,120 - \citealt{Cha15}, NGC\,7582 - \citealt{Ricci18}, NGC\,6951 - \citealt{DMay16}, NGC\,1068 - M\&S17, ESO\,428-G14 - \citealt{DMay18}). In the last three references the average misalignment is $\sim$30\textdegree, and in NGC\,7582 is nearly orthogonal. Here, we infer the torus' axis as being orthogonal to the ionization cones, misaligned by $\sim$20\textdegree~with respect to the jet.
    However, the resolved H$_{2}$ central structure has a distinct PA, with a higher degree of misalignment, of $\sim$47\textdegree. 
    Therefore, the misalignment in NGC\,4151 is still able to explain the scenario where the jet/radiative wind have ejected nuclear material in the direction of the outflow.
    
    Furthermore, the molecular walls and the presence of symmetrical bullets in the outflow indicates certain degree of organization of the nuclear gas previously settled in the vicinity of the central source. One cannot think that the same fragmentation process is occurring in both sides of the cones just because of randomly distributed gas in the ISM. Such a misalignment, between a jet and a central thick disc/torus, would naturally lead to a symmetrical interaction, which can be preserved on a larger scale. This happens because the jet/radiative wind hits the torus borders, pushing away the dense molecular gas, and creating the detected H$_{2}$ walls. Therefore, we sustain the scenario where the molecular gas is expanding from the nucleus as a molecular cavity (or cavities) and it is the material supply for most of the ionized emission seen in outflow.

    Curiously, the results found for the galaxy NGC\,6951 in \citet{DMay16} (where a jet interacts with a nuclear molecular thick disc), may indicate a good connection between molecular outflows for small and larger scales. There, the interaction is more dramatic because it involves a misaligned jet/accretion disc with respect to a thick molecular disc surrounding the nucleus (r$\sim$24 pc). In NGC\,1068 we reported signs of a similar interaction, but for scales $\sim3$ times smaller, slightly larger than the torus. Therefore, as discussed in Sect.~\ref{sec:bubble}, the misalignment between the jet/accretion disc with respect to some molecular structure (of varying scale) is a tempting scenario for the origin of some outflows and a natural explanation behind some symmetries observed along the NLR.
    We lack, however, of studies/simulations regarding the consequences of this kind of interactions.
    
    One should keep in mind that, at the scales probed here and in NGC\,1068, we are seeing the gas within the galactic disc (i.e., we cannot say that the H$_{2}$ emission, or even the outflow, are outside the disc). Eventually, in a more advanced stage of the same outflow, the H$_{2}$ emission outside the cones may be limited only to the galactic plane (more abundant in molecular material). In other words, we can only treat both the gas phases as having distinct distributions for scales larger than the galactic disc thickness. In powerful AGN, however, one may still see molecular gas in outflow at kiloparsec scales above the galactic disc \citep{Richings18a,Richings18b}.
    
    \subsection{The molecular outflow in the fountain model context}
    \label{sec:polardust}
    
    The work of \citet{Davies14} seems to suggest a common condition to detect molecular outflows if the opening angle of the ionization cones intercepts (at least in part) the galactic disc, which is rich in H$_{2}$. They identified molecular outflows in 3 out of 5 AGN, with all of them satisfying this condition. \citet{Riffel13} and \citet{Ardila04} showed that all AGN in their sample present warm nuclear molecular gas, which is associated with inflows of gas. \citet{Thaisa094151} sustain a similar interpretation for the H$_{2}$ in NGC\,4151, saying that the observed molecular gas is part of a inflowing molecular disc, partially destroyed by the radiation in the ionization cones. 
    
    Here we present a totally different scenario for the molecular gas, interpreted as outflowing material. In fact, thanks to the data treatment, we were able to detect, for the first time, a molecular outflow in this galaxy with velocities of $\sim$300 km s$^{-1}$ (Fig.~\ref{fig:H2brvmaps}, panel \textbf{d}). Moreover, this outflow seems to be spatially related to the estimated extinction inside the cones (Fig.~\ref{fig:dust}), which suggests that outflowing molecular material may be composed also by dust particles.
    The model that predicts dust and molecules in the direction of the ionization cones - the ``fountain model'', proposed by \citet{Wada12,Wada16,Wada18} - has gained major importance with the detection (through infrared interferometry) of polar dust at scales $\lesssim$1 pc (\citealt{Jaffe04,Burtscher13,Schartmann14,Tristram14,Stalevski17}).
    In the fountain model the fraction of inflowing gas that is not blown away by the accretion disc wind falls back to the same disc, which would naturally explain the thickness of the torus. However, the model predicts much more neutral atomic gas than molecules in the polar direction, with the molecular gas located almost exclusively in the inner part of the torus.  
    
    The question regarding NGC\,4151 is: would be it plausible to suppose that the detected H$_{2}$ outflow could be a consequence of the fountain model? The simulations performed by \citet{Wada16} were scaled to the case of the Circinus galaxy, with an Eddington ratio $\sim20$ times larger than NGC\,4151. For this reason, as a first approximation, one may expect that the results obtained for Circinus are an upper limit for what could be found here. 
    The distances involved for the molecular outflow in NGC\,4151 are of the order of $\sim50$ pc, 5 times more than that found in simulations. In fact, according to \citet{Wada18}, just a thick neutral atomic gas structure is found in the inner 10\,pc of Circinus. The molecular gas is limited to a thin disc, making even more difficult to explain the molecular outflow of NGC\,4151 as a consequence of the fountain model.
    
    Another possibility is that the H$_{2}$ emission is, actually, located outside the walls of the ionization cones and is seen in projection within the cones. However, the fragmentation process in both sides of the cones indicates the opposite, because the molecular bullets (clearly in outflow) are also complementary to the molecular segments from where they are located. In the SW cone, for instance, one may see only the bullets and not the extinct molecular segment \textbf{2C}. On the other hand, in the NE cone the molecules in the tenuous segment \textbf{1C} are probably being destroyed by the AGN. Moreover, the jet also seems to interact with the extremities of segments \textbf{1C} and \textbf{2B} (Fig.~\ref{fig:h2}, left panel) at the inner borders of the hourglass structure. Considering these facts, we discard the hypothesis that the molecular gas within the cones is not exposed to the central source.

       \subsection{The architecture of NGC\,4151 in light of NGC\,1068: similarities and differences}
    \label{sec:comp}
    
    In light of the new results obtained for the galaxy NGC\,1068 (M\&S17), we first compared them with respect the dichotomy found for the low- and high-velocity [Fe\,{\sc ii}] emission. We have shown that NGC\,4151 also has a low-velocity structure outlined by the walls of an hourglass geometry, representing the two ionization cones, but with a projected dimension of less than half of NGC\,1068 (155 and 380 pc, respectively). In NGC\,4151, however, this hourglass is more extended along the radio jet orientation (misaligned by $\sim$20\textdegree~with respect the ionization cones) reaching up to $\sim$190 pc.
  
\begin{figure*}
    \resizebox{\hsize}{!}{\includegraphics{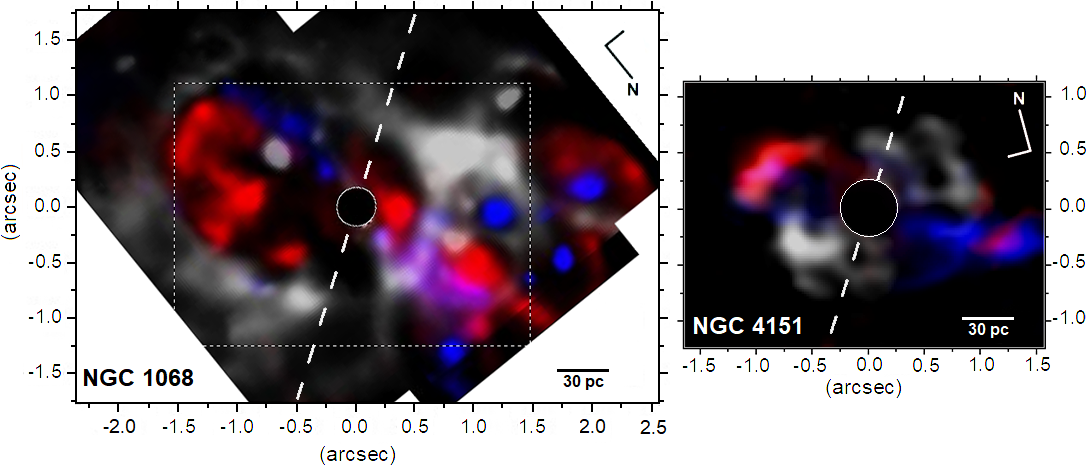}}
    \caption{High-velocity [Fe\,{\sc ii}] emission in blueshift/redshift and the H$_{2}$ molecular structure (gray) for NGC\,1068, shown in M\&S17 (left panel), and NGC\,4151 (right panel). The dashed lines are orthogonal to the cone's axis and the dashed square shows the size, in scale, of the FoV in NGC\,4151.}
     \label{fig:comp}
    \end{figure*}

    In addition, looking to Fig.~\ref{fig:comp} we see that NGC\,1068 and NGC\,4151 seem to share another common feature: a dichotomy between the cones, with an ``open'' side - when the high-velocity [Fe\,{\sc ii}] emission is seen beyond the H$_{2}$ cavity - and a ``closed'' side - when the cavity limits the high-velocity [Fe\,{\sc ii}]. An alternative definition may be given when the FoV allows to depict the entire low-velocity [Fe\,{\sc ii}] (as the case of NGC\,1068), with an open and closed side for the hourglass structure. According to the former definition, in NGC\,4151 we can identify that the NE side of the hourglass is more closed and the SW one is more open. In fact, the SW cone is possibly a transition between this dichotomy, because some [Fe\,{\sc ii}] arcs with moderate velocity are still seen crossing the interior of the hourglass (Fig.~\ref{fig:Febrvmaps}, panel \textbf{c}).
    
    Such discussion is relevant because these features are related to the side of the cone that is more fragmented. The open side in NGC\,1068, for instance, is associated with the high-velocity compact clouds already decoupled from the molecular gas. In NGC\,4151 the ionized and molecular bullets can be always associated with molecular walls, but with the SW one already destroyed by the AGN. On the other hand, in the closed side the high-velocity [Fe\,{\sc ii}] emission seems only to cover the inner part of the H$_{2}$ structure (NE cone in Fig.~\ref{fig:3f}, right panel).
    The closed cone side in NGC\,1068 also presents the high-velocity [Fe\,{\sc ii}] emission closer to the inner wall of the  molecular cavity. 
    Given the larger dimension of the outflow probed in NGC\,1068, it could have evolved to enhance bigger differences between both sides, but here the cones dichotomy already show signs that none of the sides will evolve to show a full molecular cavity. This is because we see the NE side of the cavity already being fragmented in NGC\,4151 (while the SW side of NGC\,1068 seems to be still intact).

    After confirming the same behaviour of two velocity regimes for the [Fe\,{\sc ii}] emission and the cones' dichotomy, more general questions may be raised: (1) what are the physical reasons behind these findings and (2) should we expect similar features in other AGN? We cannot probe possible answers without considering the whole scenario proposed for both galaxies, which are surprisingly similar. Starting with the low-velocity [Fe\,{\sc ii}], interpreted as the partially ionized emission from the cones' walls, this emission suggests the presence of a diffuse gas roughly distributed around the AGN, at scales of few hundreds parsecs. One possible origin of such emission could be the gas in the rotating disc detected in the H$_{2}$ and Br10 emissions, shown in Fig.~\ref{fig:rot}. However, if this was the case, we should measure some redshift at the NE wall of the hourglass and blueshift at the SW one, what is not seen. We only see the hourglass structure at the systemic velocity, although it presents an orientation somewhat close to the kinematic PA for the molecular and ionized gas. There is no hint of rotation in the low-velocity [Fe\,{\sc ii}] even at its closest emission from the nucleus (Fig.~\ref{fig:Febrvmaps}, panel \textbf{a}), where one expect to have a better match between the hourglass emitting gas and a denser region for the rotating disc. Therefore, the low-velocity [Fe\,{\sc ii}] is unlikely to be originated from an homogeneous material supply close to the nucleus.
    
    The high-velocity [Fe\,{\sc ii}], in turn, have a good correspondence with the molecular gas, like the association between the [Fe\,{\sc ii}] bullets and arcs with some H$_{2}$ segments (Sects.~\ref{sec:arc} and \ref{sec:bullets}). Since this velocity regime is closely related to the molecular walls (which is kinematically distinct from the H$_{2}$ disc; Sect.~\ref{sec:kin}), is it possible that the low-velocity one could also be related? In Fig.~\ref{fig:3f} (left panel) one may see that the most intense parts of the hourglass walls are those that seem to be spatially associated with the H$_{2}$ emission at the cones' borders. Moreover, both emissions are kinematically consistent, close to the systemic velocity. Therefore, we favour the hypothesis that both velocity regimes of [Fe\,{\sc ii}] are related to the molecular emission and the cavity expansion. In both cases the H$_{2}$ should trace the presence of dust particles (Sect.~\ref{sec:h2kin}) that, once destroyed, release the iron. In this way, the cavity expansion has a major role in the [Fe\,{\sc ii}] dynamics. Since the H$_{2}$ walls expands beyond the limits of the cones, and is pushed away within them, there will be plenty of material supply for the [Fe\,{\sc ii}] detection along the hourglass structure. Naturally, a more pronounced expansion (and fragmentation) of the molecular gas occurs into the cones, eventually extinguishing the H$_{2}$ molecules, but keeping the low- and high-velocity [Fe\,{\sc ii}] as by-product of this process.
   
    Within this scenario, the strategy of separating the [Fe\,{\sc ii}] emission in two velocity regimes gains a relative significance in characterizing outflows, in general. Such behaviour may be suggestive, for instance, of H$_{2}$ outflowing structures, normally thought as an inflowing material. Even considering that each AGN is in a distinct outflow stage, such approach can help to verify if the proposed scenario could represent a more general outflow mechanism among other AGN.
    
    The molecular gas distribution, in turn, seems to differ significantly in both galaxies: in NGC\,1068 there is a main cavity around the AGN and in NCG\,4151 the H$_{2}$ emission is found mostly outside the cones. Nevertheless, we show (in both cases) that the origin of the compact clouds (or bullets) comes from molecular walls, expanding with lower projected velocities than those found for the ionized outflow. In NGC\,1068 the cavity is described as a bubble inflated by the AGN, which is more extended to the side where no signs of fragmentation are seen. Here, the global expansion of the molecular gas has not followed the one experienced by the cones because the fragmentation is occurring at their both sides, with the SW side in a more advanced stage and possibly representing a former molecular wall. The current connection between the molecular and ionized gas phases, as well as the direction of the outflow, lead us to the same scenario found for NGC\,1068, but with a more complex structure of bubbles/cavities.

    Moreover, the close relation between the radio knots with the molecular walls and the [Fe\,{\sc ii}] bullets in NGC\,4151 indicates that the jet plays an important role in shaping the NLR.
    NGC\,1068, however, has an additional mechanism responsible to fragment one side of the molecular wall (open cone), which is the production of a thermal wind (called ``secondary wind'') launched from the spot where the jet bends and injects its energy into a molecular cloud, close to the nucleus. Such dramatic interaction is not seen in NGC\,4151, but a weak low-velocity [Fe\,{\sc ii}] emission outside the SW part of the hourglass (structures \textbf{S1} and \textbf{S2} seen in Fig.~\ref{fig:FeII}) may be originate from the jet/ISM interaction where two radio knots (\textbf{C2} and\textbf{C3}) are associated with molecular walls. These spots might represent the location of a smaller scale thermal secondary wind, with the bullet B-si4 (Fig.~\ref{fig:bullets}) possibly ionized by this wind.     
    
    With respect to the coronal emission, both galaxies also show strong similarities, as in the origin of the highly ionized compact clouds and the low-velocity emission associated with the jet. The difference is that NGC\,4151 presents two distinct orientations for the two velocity regimes, indicating that the fragmentation process does not depend, at least entirely, on the jet. However, since the molecular gas is expanding, it is quite likely that the jet already interacted strongly with other parts of the current molecular structure, in an earlier stage.  
    
    The apparent lack of an association between the high-velocity CL and the soft X-ray emission in NGC\,4151 is intriguing, which does not occur for NGC\,1068.
    In the galaxy ESO\,428-G14 the bubble seen from the coronal emission is not totally associated with the X-ray emission nor to the jet (just only half of its inner wall). In this case, one side of the bubble is possibly shock-excited by the jet and the other is not. In NGC\,4151, there are two possible explanations for these highly ionized bullets (not directly associated with the jet): 1) they could represent a less dense gas able to be photoionized by the AGN or 2) they could have interact with the jet in the past and are still emitting. Giving the scenario of the expanding bubble, we favour the hypothesis of the jet shock-excitation (what would take into account the bullet B-si4, not exposed to the AGN).
    
    Looking to the X-ray spectra of both galaxies, they are surprisingly similar (\citealt{Kink02,Schurch04}), and it is hard to distinguish between their line ratios and even the FWHM of their emissions. For this reason, \citet{Schurch04} modelled the NLR of NGC\,4151 with the same geometry adopted for NGC\,1068 in \citet{Kink02}, resulting in a good fit for both cases. The only appreciable difference is the ionic density, which is lower in NGC\,4151. The authors favour the hypothesis that the NLR gas in NGC\,4151 is denser than in NGC\,1068.

    One should keep in mind that we are comparing a Seyfert 1 galaxy (NGC\,4151) with low Eddington ratio ($L_{Bol}/L_{Edd}\sim$0.013 - \citealt{Birzan08,Bentz06}), with a Seyfert 2 galaxy (NGC\,1068) accreting gas at a much higher rate ($L_{Bol}/L_{Edd}\sim$0.44 \citealt{Pier94,Greenhill97}, or even higher - 0.77, \citealt{Lodato03}). In other words, there is support for a radiation pressure 30 to 60 times higher in NGC\,1068. From this fact, we may expect a larger relative dimension for the outflow in NGC\,1068, as detected. With a low Eddington ratio and a jet power $\sim10^{5}$ times weaker than NGC\,1068 \citep{Ulvestad05} it is surprising that both galaxies present so many similarities between each other.
    
    With a similar process of molecular gas fragmentation in both galaxies, we can safely say that the outflow in NGC\,4151 represents a more compact version than the one seen for NGC\,1068. The ``age'' of the detected outflow can be estimated from the expansion rate of the molecular cavities, corrected from projection effects.
    If we assume a constant expansion velocity of 260 \kms for the molecular gas in the NE ``closed'' cone, at a distance of 60\,pc, and an inclination i=45\textdegree~for the bi-cone \citep{Das05}, we find a time scale of 1.2$\times10^{5}yr$ for the outflow in NGC\,4151. Although rather simplistic, the same estimate applied to NGC\,1068 (assuming the maximum H$_{2}$ velocity detected for the cavity, of $\sim$300 \kms; Fig.20 in M\&S17) results in a time scale of 2.9$\times10^{5}yr$. Within the involved approximations, both outflows are nearly contemporaneous, with the outflow of NGC\,4151 being driven by a less powerful AGN (at least in the current state).
    
    According to \citet{Wang11b}, the estimated age for the interaction between the jet and the hot gas (at the same scales) is $\sim10^{5}$\,yr, in agreement with our outflow timescale. Furthermore, the extended ($\sim$2\,kpc) soft X-ray emission requires a timescale of the same order for the deposit of mechanical energy from the central source \citep{Wang10}. These authors also argue that, for the gas to be photoionized by the AGN, it implies in a last episode of high activity phase occurred $\lesssim2.5\times10^{4}$\,yr ago.
    According to our estimated age, this means that the AGN could have been brighter $\sim10^{5}$\,yr ago and maintained this high activity phase until fading to the current stage in the last $\lesssim2.5\times10^{4}$\,yr.
    If correct, such time scales mean that 80 per cent of the gas expansion time was carried out under a more powerful AGN. Such fact suggests that both AGN could have shared similar Eddington ratios in the past, leading more naturally to the current similarities found for both outflows.

    \section{Conclusions}
    \label{sec:conclusions}
    
    We have analyzed archival data of NGC\,4151 in the $H$ and $K$ bands with AO and employed a set of image processing techniques developed by our group. The spatial resolution of the Br10 and Br$\gamma$ images after the data treatment is comparable to the H$\alpha$ image of HST. Our main conclusions are listed bellow.
    
    \begin{enumerate}
    
    \item The low-velocity [Fe\,{\sc ii}] $\lambda$16440 \AA~emission (corresponding to the velocity range between -92 km s$^{-1}<v<73$~km s$^{-1}$) depicts the walls of an hourglass geometry, with a PA=58\textdegree$\pm3$\textdegree, with the SW lobe more extended than the NE one. This orientation implies a dusty torus with PA=148\textdegree~and the hourglass geometry is similar to the one found for NGC\,1068.\\
    
    \item The high-velocity [Fe\,{\sc ii}] (corresponding to the velocity ranges of -1732 km s$^{-1}<v<-383$~km s$^{-1}$~and 401 km s$^{-1}<v<1276$~km s$^{-1}$) present redshift and blueshift in both sides of the ionization cones, with a preference for blueshift in the SW cone and redshift in the NE one. This emission is contained inside the low-velocity [Fe\,{\sc ii}] emission (the hourglass walls) and seems to be more fragmented, like the case of NGC\,1068.   \\
    
    \item In contrast to the [Fe\,{\sc ii}], the Br$\gamma$ and Br10 emissions present almost no emission along the hourglass walls, being more limited to its interior. This suggests that the low-velocity [Fe\,{\sc ii}] comes from the partially ionized region associated with the walls of the hourglass/ionization cones. \\
    
    \item Most of the H$_{2}$ emission is distributed at the shadow of the torus, with some emission peaks close to the hourglass walls and, possibly, to the radio jet. Moreover, the low-velocity H$_{2}$ structure has an estimated orientation $\sim$27\textdegree~less than the expected for a collimating structure.   \\
    
    \item We have detected a molecular outflow for the first time in NGC\,4151, which is within the ionization cones and have velocities of $\sim$300 km s$^{-1}$. This faint high-velocity emission is confirmed both for the $H$ and $K$-bands.   \\
     
   \item We have found 16 compact clouds of gas, or ``bullets'', indicating bubbles of ionized and molecular gas. Six pairs of them preserve some degree of symmetry in distance and velocity with respect to the nucleus. These structures seem to represent the fragmentation of the molecular walls in which they are located.   \\
    
	\item Unexpectedly, the X-ray structures depicted by the [Ne\,{\sc ix}] emission are unrelated to the overall [Si\,{\sc vii}] coronal line emission, but are surprisingly similar to the [Fe\,{\sc ii}] high-velocity emission, with the same estimated de-projected velocities of $\sim$650 \kms. 	\\
	
	 \item We detected a transient and possibly unresolved ($\lesssim$10 pc) emission in $\lambda$16525 \AA, which is interpreted as a redshifted [Fe\,{\sc ii}] emission with v=1550 km s$^{-1}$ due to Doppler effect. Such emission is possibly a response to the variability previously detected for the broad lines in this galaxy and would be located right above the BLR. \\
	 
	 \item We change the paradigm of the misalignment between the jet and the ionization cones for this galaxy, showing that it is not a projection effect. As a consequence, the accretion disc is misaligned with respect to the torus by, at least, $\sim$20\textdegree. \\
	 
	 \item We show that the jet has a clear influence in shaping a fraction of the NLR, with the double peaked narrow lines in the vicinity of the radio knots \textbf{C2} and \textbf{C5} possibly representing the lateral expansion of the gas. Close to these knots we see more signs of a strong jet-cloud interaction, as the highest detected velocities for the [Fe\,{\sc ii}] bullets and the highest [Ne\,{\sc ix}]/[O\,{\sc vii}] ratios, both associated with molecular walls. \\
	 
	 \item Inspired by NGC\,1068, a new architecture is proposed for NGC\,4151: the ionized outflow would be mostly a consequence of a molecular fragmentation process, leading to the formation of bullets of gas. We suggest that the low-velocity H$_{2}$ structures outside the cones, which are connected to the [Fe\,{\sc ii}] emission and all the detected bullets, are also in outflow. Such facts support the scenario where an expanding molecular bubble (naturally seen in low-velocity when projected in the plane of the sky) is being inflated and disrupted by the AGN and, possibly, also by the radio jet.

		\end{enumerate}

    \section*{Acknowledgments}

     We thank the referee for carefully reading the manuscript and making very constructive comments for the clarity of the paper.
     This work was supported by FAPESP (Fundaç\~ao de Amparo \`a Pesquisa do Estado de S\~ao Paulo), under grants 2011/19824-8 (DMN) and 2011/51680-6 (JES), and by the Oxford Centre for Astro-physical Surveys, which is funded through generous support from the Hintze Family Charitable Foundation. We also acknowledge Jodrell Bank Centre for Astrophysics, which is funded by the STFC, and Michele Cappellari for the IDL implementation of the PA routine. The radio observation for this research was obtained with e-MERLIN (formerly MERLIN), a National Facility operated by the University of Manchester at Jodrell Bank Observatory on behalf of STFC. The radio data was kindly provided by Rob Beswick and Ian McHardy, who were the PIs on LeMMINGs that allowed its distribution. JW acknowledges support by the National Key R\&D Program of China (2016YFA0400702) and the National Science Foundation of China (U1831205). The author also appreciate the thoughtful revision carried out by Yaherlyn Diaz at Universidad de Valparaíso, Chile.

    \bibliographystyle{mnras}
    \bibliography{biblio}


    \appendix
    
    \section{A transient [Fe\,{\sc ii}] emission}
    \label{sec:tr}
    
    Looking to the $H$-band nuclear spectra in Fig.~\ref{fig:spec} one may note a strong nuclear emission line $\sim$80\AA~to the red from the [Fe\,{\sc ii}] $\lambda$16440 \AA~line. Before discussing its possible origins we have first to confirm if this feature is, in fact, real. Since this emission is present in all of the 9 data cubes used in our analysis, we extracted its spatial and spectral profiles from the combined data. The measured FWHM for both spatial directions is 0.182$\pm$0.011 and 0.143$\pm$0.005 arcsec for the $x$ and $y$-axis, respectively (only slightly larger than the values measured for the PSF extracted from the standard star used in the deconvolution process). The line profile, in turn, has a FWHM of $9\pm2\AA$, $\sim3$ times the instrumental resolution. From this result we concluded that this emission went through the atmosphere and should be an unresolved (or almost unresolved) emission from the AGN. This emission, however, is not seen in the spectra presented by \citet{Thaisa094151}, observed $\sim$1 yr before, what gives its transient nature.  
    
     We favour the hypothesis that this transient emission represents the [Fe\,{\sc ii}] $\lambda=$16440\AA~with an unusual redshift and brightness. To test this hypothesis we should see this emission in other [Fe\,{\sc ii}] lines as well, thus, with this aim, we have applied PCA tomography \citep{Steiner09} and found an anti-correlation between the [Fe\,{\sc ii}] $\lambda=$16533\AA~and 16440\AA~lines with respect to their transient emissions in redshift. Their inferred velocities are 1760$\pm$108 and 1550$\pm$91 \kms, respectively, which are compatible with each other within their uncertainties. The measured flux for the transient redshifted [Fe\,{\sc ii}] $\lambda=$16440\AA~line is 4.58 $\pm$0.33 $10^{-15}$~erg s$^{-1}$~cm$^{-2}$ for a nuclear aperture radius of 0.25 arcsec, which represents 24 per cent of the non-redshifted emission.  
   
    A study concerning the temporal variation of the line profile in some IR emissions was performed by \citet{Esser19} in NGC\,4151, which noted an increasing line flux ($\lesssim2$ times) for a velocity interval between 2000 and 3000 km s$^{-1}$, in redshift. This variation was detected along a period of 3 years, with a peak around 2006. Such behaviour coincides with a smaller torus radius measured by \citet{Koshida09}, from where it is suggested that the origin of this high-velocity emission comes from the dust sublimation. 
    However, the transient [Fe\,{\sc ii}] emission is detected only $\lesssim$1 year later. If one assume a cause and effect relationship between a phenomenon located in the BLR and the redshifted [Fe\,{\sc ii}] emissions, it would result in an upper limit of 1 light-year for some physical response, roughly outside the BLR.

    The second hypothesis is that this emission could come from a gravitational redshift, in which the distance from the SMBH can be evaluated from the following equation:

    \begin{equation}
    \begin{aligned}
    1 + z_{grav}&=\left(1 - \frac{2GM_{BH}}{rc^{2}}\right)^{-1/2}
    \label{redshift}
    \end{aligned}
    \end{equation}
    
    \noindent where $z_{grav}$ can be approximated as $v/c$, $v=1655$ \kms (average value from the two transient Fe\,[{\sc ii}] lines), $G$ is the gravitational constant and $M_{BH}=4.57\times10^{7}$ \Msun~\citep{Bentz06}. 
    The result, however, rules out the hypothesis of gravitational redshift, because the adopted velocity corresponds to a distance of only $\sim0.5$ light-days, or 90 Schwartzchild radius. In fact, this radius would be $\sim20$ times smaller than the size of the accretion disc expected for this AGN \citep{Morgan10}. Furthermore, a prohibited emission would be inconceivable for such small volumes and high densities, when compared to the NLR.
    Therefore, we maintain the hypothesis that this emission represents an event located at the basis of the NLR and could be possibly a response to the high-velocity line profile variations detected by \citet{Esser19}.
    
     \section{Stellar kinematics}
    \label{sec:sk}

    \begin{figure}
    \resizebox{0.95\hsize}{!}{\includegraphics{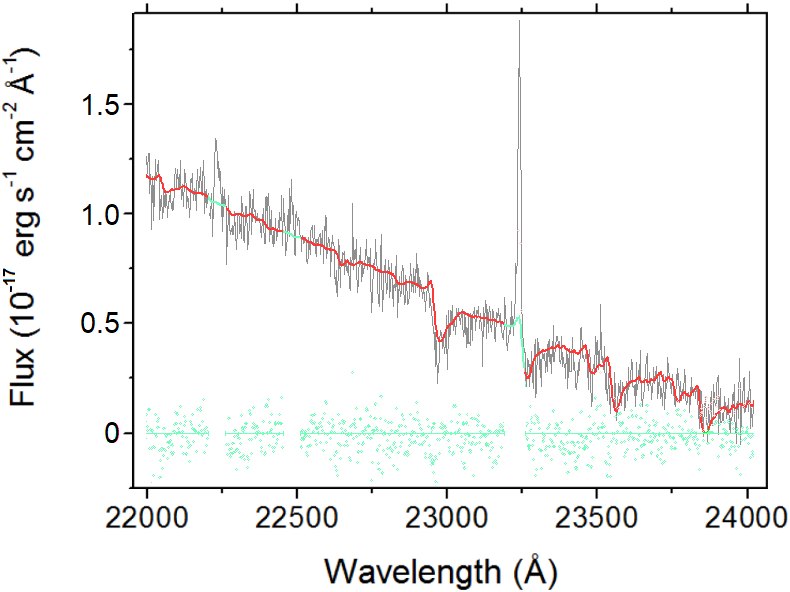}}
    \caption{Spectrum form the $K$-band at 0.25 arcsec NW from the nucleus with the resulting fit using pPXF (red). The emission lines were masked and the residuals are shown in green.}
     \label{fig:ppxffit}
    \end{figure}
    
    \begin{figure}
    \resizebox{0.95\hsize}{!}{\includegraphics{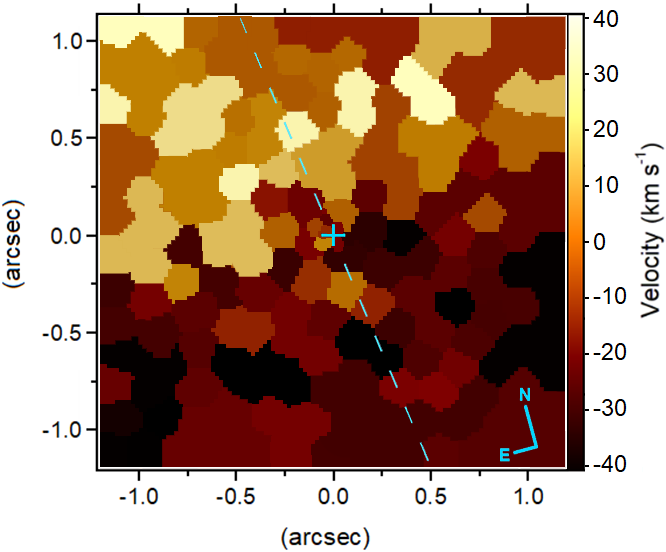}}
    \caption{Stellar velocity map with bins of equal S/N. The dashed line denotes the kinematic PA, of 8\textdegree$\pm$3\textdegree.}
     \label{fig:ppxfvel}
    \end{figure}

    We measured the stellar stellar kinematics with the penalized  pixel fitting (pPXF) method \citep{Cappellari04} (see the example of a fitted spectrum in Fig.~\ref{fig:ppxffit}), resulting in a PA of 8\textdegree$\pm$3\textdegree~for the central 1 arcsec (Fig.~\ref{fig:ppxfvel}), $\sim$10\textdegree~smaller than the one found by \citet{Dumas07} (19\textdegree$\pm$2\textdegree) for a FoV 10 times larger. On the other hand, a larger set of the same observations used here provided a PA similar to the one found by the later authors, for a FoV of 1.5 arcsec$^{2}$ \citep{Onken14}. We noticed, however, that some bins to the NW side of their kinematic map (Fig.3, left panel) favour a smaller PA for a smaller FoV, which is our case. Given our uncertainty of only 2\textdegree, we believe that this difference is real and the very central stellar kinematic has a twist of $\sim$10\textdegree to a smaller PA. 
    
    The velocity ranges from -40 to 40 \kms, which agrees with the values found by \citet{Onken14} and, according to \citet{Dumas07}, rises to -70 to 70 \kms for a spatial scale of 20 arcsec.
    The work of \citet{Dumas07} found a difference of $\sim$20\textdegree~between the PA for the stars and the gas, indicating that they rotate in distinct planes. 
    From these measurements we see that our PA, for a smaller FoV (both for stars and the gas), are compatible with the extended galactic kinematics, with the gas presenting disc-like rotation for a PA similar to that we have found.

    \label{lastpage}

    \end{document}